\documentclass[a4paper, 12pt]{article}
\usepackage{comment}

\usepackage{amsmath, amssymb, amsthm}
\usepackage{graphicx,color}
\usepackage{graphics}
\usepackage{rotating}

\usepackage[utf8]{inputenc}

\usepackage[T1]{fontenc} 

\usepackage{lmodern} 

\usepackage{natbib} 
\setcitestyle{authoryear,open={(},close={)}} 

\usepackage[english]{babel}

\usepackage{booktabs}
\usepackage{amsmath,amsthm}

\usepackage{todonotes}
\usepackage{setspace}   
\usepackage[pdfborder={0 0 0}]{hyperref} 
\usepackage{threeparttable} 
\usepackage{tablefootnote}

\usepackage{booktabs}   

\usepackage{dcolumn}    
\usepackage{caption}    
\usepackage{adjustbox}  

\usepackage{placeins}   
\usepackage{float}      
\usepackage{authblk}    

\usepackage[margin=1in]{geometry}
\usepackage{lineno} 

\usepackage[toc,page]{appendix} 
\usepackage{subcaption} 
\usepackage{indentfirst}

\hypersetup{
colorlinks,
citecolor=blue,
linkcolor=blue,
urlcolor=blue} 
\graphicspath{{figures/}}
\title{A Political Economy Definition of the Middle Class}
\author{Alejandro Corvalan}
\date{December 2025}

\begin{document}

\maketitle
\doublespacing

\begin{abstract}

Economists often define the middle class based on income distribution, yet selecting which segment constitutes the `middle' is essentially arbitrary. This paper proposes a definition of the middle class based solely on the properties of income distribution. It argues that for a collection of unequal societies, the poor and rich extremes of the distribution unambiguously worsen or improve their respective income shares with inequality. In contrast, such an effect is moderated at the center. I define the middle class as the segment of the income distribution whose income shares are insensitive to changes in inequality. This unresponsiveness property allows one to single out, endogenously and with minimal arbitrariness, the location of the middle class. The paper first provides a theoretical argument for the existence of such a group. It then uses detailed percentile data from the World Income Database (WID) to empirically characterize the world middle class: a group skewed toward the upper part of the distribution—comprising much of the affluent population below the very rich—with stable borders over time and across countries. The definition aligns with the prevailing view in political economy of the middle class as as a moderating actor, given their null incentives to engage in distributive conflict.

\end{abstract}

\newpage

\section{Introduction}

The presence of a large middle class is considered an important determinant of democracy, social stability, and economic growth. The literature in economics has discussed the generally beneficial role of the middle class on industrialization \citep{murphy1989income}, entrepreneurship \citep{acemoglu1997prometheus}, human capital \citep{galor1997technological}, democracy \citep{barro1999determinants} and growth \citep{easterly2001middle}. As \cite{banerjee2008middle} summarize, “we expect a lot from the middle class”.

However, there is much less consensus when it comes to defining the middle class. Economists often characterize the middle class based on income distribution, yet selecting which segment constitutes the 'middle' is largely discretionary. The middle class has been characterized in terms of income or percentiles, but different lower and upper bounds are used depending on the question and countries \citep{ravallion2010developing, gornick2014income}. This ambiguity is not innocuous, given that empirical results will depend on how the middle group is defined \citep{atkinson2013identification, ricci2020measure}.

This paper proposes a definition of the middle class based solely on the properties of income distribution and inequality. It argues that for a collection of unequal societies, the poor and rich extremes of the distribution unambiguously worsen or improve their respective income shares with inequality. In contrast, such an effect is moderated at the center. I define the middle class as the segment of the income distribution whose income shares are insensitive to changes in inequality. This unresponsiveness property allows one to single out, endogenously and with minimal arbitrariness, the location of the middle class.

A mechanical analogy is illustrative. The income distribution behaves like a seesaw when inequality changes. At the two ends of the board sit the poor and the rich: when inequality increases, the poor end goes down while the wealthy end rises; when it decreases, the movement reverses. All of this motion occurs around the fulcrum, or pivot point — the central support that allows the seesaw to rotate. That fulcrum is the middle class. All income distribution percentiles pivot around a fixed middle class when inequality changes.

That the middle of the income distribution is remarkably stable across countries and over time is not a novel observation, as documented by \cite{deininger1996new, deininger1998new}. Similarly, Palma identifies a group of deciles displaying an exceptionally low coefficient of variation over a sample of countries, to the point of treating them as effectively constant \citep{palma2006globalizing, palma2011homogeneous, palma2019behind}. By defining the middle class as this invariant segment, inequality is reduced to the income ratio between the rich and the poor. However, this same simplification precludes a meaningful conception of the middle class that varies across countries. In addition, insensitivity to inequality is central to the conceptual interpretation of the middle class.

The definition of the middle class as non-responsive to distributive changes is in line with the idea of moderation that runs through the political economy literature. The notion of the middle class as a stabilizing force can be traced back to Aristotle, who viewed it as a virtuous group that tempers the excesses of the rich and the poor \citep{AristotlePolitics}. Similarly, \cite{lipset1959some} argues that the middle class lacks strong material incentives for conflict, while \cite{benhabib2006political} model a setting in which the rich or the poor — though not the middle — may turn against democracy. The middle class identified in this paper corresponds precisely to such a group: being insensitive to changes in inequality, it has no stake in distributive conflict. Since conflict is costly, the invariant middle class primarily prefers to avoid it, thus exhibiting a clear preference for moderation.

The paper proceeds as follows. The first part defines and discusses the existence of an endogenous middle class. The definition rests on two conditions. The first pertains to the size $M$ of the middle class, which is arbitrary. The second relates to its insensitivity to inequality, a robust statistical feature of income distributions: across societies with different levels of inequality, a range of central percentiles have income shares that remain essentially unchanged as inequality varies. Under general assumptions about the extremes and a continuity argument, I demonstrate the existence of a class that satisfies both conditions.

I characterize the income distribution using a simple, flexible Pareto parametrization to test the previous theoretical arguments. The simulated distributions replicate the formal results, revealing the existence of a middle class that is unresponsive to inequality. Interestingly, the exercise shows that this middle group lies far from the median and is instead located closer to the richer upper end of the distribution.

Secondly, I compute the world middle class using data from the World Inequality Database (WID). WID provides detailed income-share data for all percentiles across 215 countries since 1980, comprising nearly one million data points. I use the full country–year sample to identify the middle class. For size $M=50$, the middle class corresponds to the percentile interval $(48, 98)$; for $M=30$, it corresponds to $(65, 95)$. These intervals are robust along several dimensions. Middle-class income shares are strongly correlated across alternative size definitions. Moreover, the middle's boundaries are consistent with several inequality measures. Examining the middle class over time, I show that it has been remarkably stable, remaining essentially unchanged from 1980 to the present. When comparing countries, most show middle-group intervals that are very close to the global benchmark. Overall, the middle class is highly consistent in both the between- and within-country sample analyses. 

The immediate observation about the inequality unresponsive middle class is that it is not close to the median income of the distribution but skewed toward the richer percentiles. It includes all the affluent segments except for the very rich. While some literature locates the middle class near the median, several works instead associate it with relatively well-off groups just below the top \citep{milanovic2002decomposing, birdsall2010indispensable, palma2011homogeneous, piketty2014capital}. A relatively wealthy middle class is also consistent with this group’s self-perception, as several surveys document\citep{pressman2007decline, eisenhauer2008economic, cashell2008middle}. In addition, because the definition of the unresponsive middle class is based on fixed percentiles, its variance is determined by the income shares held by those percentiles. I describe those shares and their global evolution from 1980 to the present.

The final section illustrates how empirical research can benefit from the information embedded in this measure. From a political-economy perspective, a middle class that is insensitive to inequality has no material stake in distributive conflict and can therefore act as a moderating force. Accordingly, the analysis first examines the association between the middle class and democracy. I regress measures of democracy on the size of the middle class and show that the lagged income share of this group—defined as in this paper—exhibits a positive, statistically significant, and robust coefficient across all specifications. These estimates are not intended to validate the proposed interpretation of the middle class, but rather to illustrate the measure's empirical relevance and practical usefulness.

\section{Who is Middle Class?}

The dominant approach in economics for measuring the middle class is income-based, meaning that the definition is based purely on personal income. On the one hand, income is typically seen as a good proxy for living standards and welfare, strongly correlated with education, health, housing, and other measures linked to class. On the other hand, there are reliable and detailed statistics on income.\footnote{Today, datasets such as WID and LIS provide detailed information about income across the entire distribution.} Economists are aware that income-based middle classes are not perfectly comparable with the more multidimensional approach to classes in other disciplines. \cite{atkinson2013identification} argued that the concept of class requires the examination of other dimensions beyond income, and they mentioned property and occupation as leading alternatives.\footnote{Relatedly, \cite{edo2021multidimensional} proposes a multidimensional identification of the middle class that extends beyond income.} 
 
Crucially, income-based definitions of the middle class are discretionary.\footnote{An exception is \cite{foster2010polarization}. However, their definition requires observing the distribution over time to study their polarization dynamics.} With no information other than "being in the middle", any notion of the middle class is subject to a high level of arbitrariness. The literature reports different definitions depending on purpose and across countries at different stages of social and economic development  \citep{gornick2014income, ravallion2010developing}.\footnote{ \cite{reeves2018dozen}, for instance, listed a dozen ways to measure the middle class.} This ambiguity in the definition of the middle class is not innocuous, given that empirical results might change depending on the particular definition \citep{atkinson2013identification}. 

The following paragraphs summarize the approaches for an income-based definition of the middle class. The most widely used individual or household income measure is disposable income, which considers the payment of direct taxes. For an extended review, see \cite{ricci2020measure}.

\subsection{Definitions based on Absolute and Relative Income}

First, several researchers define the middle class in terms of absolute income. In addition to their direct interpretation, income values allow comparison of the lower limit of the middle class with the poverty line. For instance, the fixed-income approach has been used to estimate the so-called global middle class. \cite{milanovic2002decomposing} considered individuals living with a per capita income of \$12-50 a day, in 2000 purchasing power parity terms (PPP), as the middle class in the world. A similar approach is \cite{kharas2010new}, which chooses a range of daily expenditures of \$10 and \$100 in 2005 PPP terms.

Fixed-income thresholds are also used for comparative analysis. Because income levels are not directly comparable across countries at different stages of development, different studies apply different thresholds. \cite{banerjee2008middle} consider the middle class in developing countries as households with per capita daily consumption between \$2 and \$10 (PPP). In contrast, \cite{ravallion2010developing} defines the developing world middle class as the income range between the median poverty line of 70 countries (\$2 per day at 2005 PPPs) and the US poverty line (\$13 a day at 2005 PPPs). \cite{lopez2014vulnerability} defined the middle class in Latin America and the Caribbean (LAC) based on the notion of economic security, ranging from 10\$ to 50\$ in 2005 PPPs per day.

A second definition uses relative income to identify the middle class. The approach uses the income threshold as a percentage of the distribution's median income. A first convention identifies the middle class as households with incomes between 75 and 125 per cent of the national median \citep{thurow1987surge, pressman2007decline}. However, those values are symmetric, while the income distribution is skewed to the right, resulting in many more individuals with incomes below the mean. Accordingly, other authors prefer asymmetric bounds, with the upper cut-off in 150, 200, and even 250\% of the median income, as reported by \cite{ricci2020measure}. 

There have been attempts to justify the discretionary election of the bounds. The literature relates the lower cut-off to the poverty line, often about 60\% of the mean income, and the choice of 75\% provides an additional margin above \citep{thurow1987surge,atkinson2013identification}. It is more challenging to provide a rationale for setting the upper margin and, in many cases, the choice arises from the inspection of the same data. \cite{birdsall2010indispensable} proposes a hybrid definition, while \cite{eisenhauer2008economic} uses additional information about wealth to define the upper bound. 

International institutions' choice of the middle class illustrates the discretion in all the above definitions. The World Bank prefers a fixed-income measure, with households with daily income between \$10 and \$50 per person (PPP) as the middle class. In contrast, the OECD defined the middle class as households with incomes between 75\% and 200\% of median income. 

\subsection{Definitions based on Percentiles}

An alternative approach identifies the middle class using income distribution percentiles. As percentile data is recent, definitions typically rely on deciles or quintiles. 

The first approach is to define arbitrary percentiles close to the median. As for absolute and relative income, the election of the upper and lower percentiles bounds is arbitrary. The preferred middle class is the middle three quintiles of the family income distribution, such as in \cite{levy1987middle,barro1999determinants}. \cite{easterly2001middle} identifies the “middle class” as those between the 20th and 80th percentile on the consumption distribution. \cite{alesina1996income} and \cite{deininger1996new} chose a middle class composed of the third and fourth deciles. 

In the case of percentiles, using the size of the middle class as a measure is pointless. Therefore, the analysis of the middle class considers their income share across countries and over time. The advantage of percentiles is that they are comparable across countries and epochs, as emphasized in the top 1 percent literature \citep{piketty2006evolution, piketty2014capital}. The disadvantage is that the measure is invariant to spread \citep{foster2010polarization}. 

A different approach to define a middle class based on percentiles is \cite{palma2006globalizing, palma2011homogeneous, palma2019behind}. The author noted the following empirical regularity: the individuals in deciles 5 and 9, ranked between 0.40 and 0.90 in the income distribution, own about $50\%$ of the national income. Using this result, the author concludes that the variance of cross-country inequality lies in the remaining deciles. The Palma middle class is the supposedly invariant middle $(40,90)$.

Palma's definition privileges invariance per se, measured, for instance, by the coefficient of variation. The rationale for this invariance is to summarize the income distribution by the two non-invariant extremes: the rich and the poor. The empirical regularity noted by Palma - that the middle of the income distribution is homogenous and the role of inequality is thus only noted in the tails - is central for this work. However, I depart from Palma's by recognizing that the middle class is not invariant but rather insensitive to inequality. The difference is fundamental for interpretation, since it suggests a class less prone to distributive conflict. However, more importantly, it does not assume that the middle class is fixed, thus allowing the construction of a middle class with country- and year-specific variation, as measured by income shares.

The Palma middle class, along with the empirical observation that supports it, has received criticisms. The first is that $(40,90)$ is arbitrary, and there could be other intervals where the variance is even smaller. For instance, \cite{krozer2015inequality} proposes a more invariant middle class as $(40,95)$. This arbitrariness and the fact that it depends on the same distribution and thus changes over time were criticized by \cite{milanovic10still}. In addition, \citep{cobham2016inequality} shows that the Palma index highly correlates with inequality, so it is not completely fixed. In short, the observed empirical regularity does not lead to a strictly invariant, non-arbitrary, and robust definition of the middle class.

\section{The Middle Class: A Definition}

This section endogenously defines the middle class. The definition builds on the link between income shares and inequality. To study such a relationship, a sample of country-year observations is required. For simplicity, I refer to these units as ‘countries’ throughout the text, although the sample can be a cross-section, a panel of country-year units, or the same country at different times. 

The definition of the middle class is thus sample-dependent. This dependence is an essential requirement of the formulation, given that a unique distribution is not enough to establish the relationship between the different parts of the distribution and inequality. Yet it seems strange that inequality movements in one country can potentially change the middle class in another. The underlying assumption is that the relationship between shares and inequality has a specific structure, and countries over time are realizations of the process.\footnote{Variable definitions based on parameters estimated from a regression on the full sample appear quite often in economics and finance. A well-known example is risk-adjusted returns, which first compute returns by estimating a Fama-French model across all assets, and then use the estimated parameters to define the abnormal return of a particular asset. Similarly, the factor productivity of a firm requires a previous regression of capital and labor on output production for the entire sample of firms.}

I further assume that, for an arbitrary sample, the relationship between inequality and income shares is linear. A linear specification simplifies the analysis and provides a straightforward argument for the existence of a middle class. However, linearity is a strong assumption, and I will discuss it with some detail at the end of the section.


\subsection{Definition}

Let $S(p,q)$ be the income share between the quantiles $p$ and $q$, with $0 \leq p < q \leq 1$. Notice that $S(0,1)=1$. I assume that differences in inequality across countries mostly explain the variances in income shares. For a country $i$, with Gini index given by $G_{i}$,\footnote{Admittedly, the definition depends on the inequality index we choose, but we will show in the empirical application that the results are robust to the use of several indices. As Gini is the most used index for cross-country comparisons, I use Gini as the preferred inequality measure.} income shares are linear functions of inequality: 

\begin{equation}
S(p,q)_{i}=\alpha(p,q)+\beta(p,q)G_{i}+\varepsilon_{i}
\label{eqn:linear}
\end{equation}

with $\varepsilon_{i}$ an error term satisfying the standard exogeneity condition $E(\varepsilon|G)=0$. I assume that $\alpha$ and $\beta$ are continuous in their arguments $p$ and $q$. The continuity assumption rules out discrete jumps in the income distribution. 

\textbf{DEFINITION.}The middle class is a quantile interval $(p,q)$ of size $M$ that is non-responsive with respect to $G$. Formally, it satisfies the following two conditions:

\begin{subequations}
\begin{eqnarray}
\beta(p,q)=0 \label{eqn:def1} \\ 
q-p=M \label{eqn:def2}
\end{eqnarray}
\end{subequations}

The first condition (\ref{eqn:def1}) implies that the middle class does not correlate with inequality. There are many pairs $(p,q)$ that satisfy this condition, so the second one, the size of the middle class (\ref{eqn:def2}), is needed to pin down a particular subgroup. The size $M$ is arbitrary.

\subsection{Existence}

To provide an intuition about the existence of the middle class, we assume the following behavior of the coefficient beta at the borders. Consider $0<\varepsilon\ll 1$ such that:

\begin{subequations}
\begin{eqnarray}
\beta(0,\varepsilon)<0 \label{eqn:ineq1} \\ 
\beta(1-\varepsilon,1)>0  \label{eqn:ineq2}
\end{eqnarray}
\end{subequations}

The inequalities indicate that the lowest quantile of the income distribution, for instance, the first percentile, decreases its income share as the Gini increases. The opposite is also true, meaning that the richest segment of the distribution, let us say the 100th percentile, increases its income with the Gini.

These relationships are assumptions, given that inequality depends on the entire distribution, and it is easy to construct examples in which these expressions do not hold. However, they are consistent with the Pigou-Dalton Principle of Transfers. According to the principle, any transfer from a lower to an upper percentile increases inequality. So any reduction of the quantile $(0,\varepsilon)$ due to transfer will increase inequality, consistent with equation (\ref{eqn:linear}) and (\ref{eqn:ineq1}). A similar argument applies to the richest percentile. Empirically, the literature on poverty, on the one hand \citep{nolan2011economic}, and on top incomes, on the other \citep{piketty2006evolution}, confirms these inequalities.

\begin{figure}[H]
    \centering
\centering
  \caption{Existence of the  Middle-Class}
  \includegraphics[width=0.8 \textwidth]{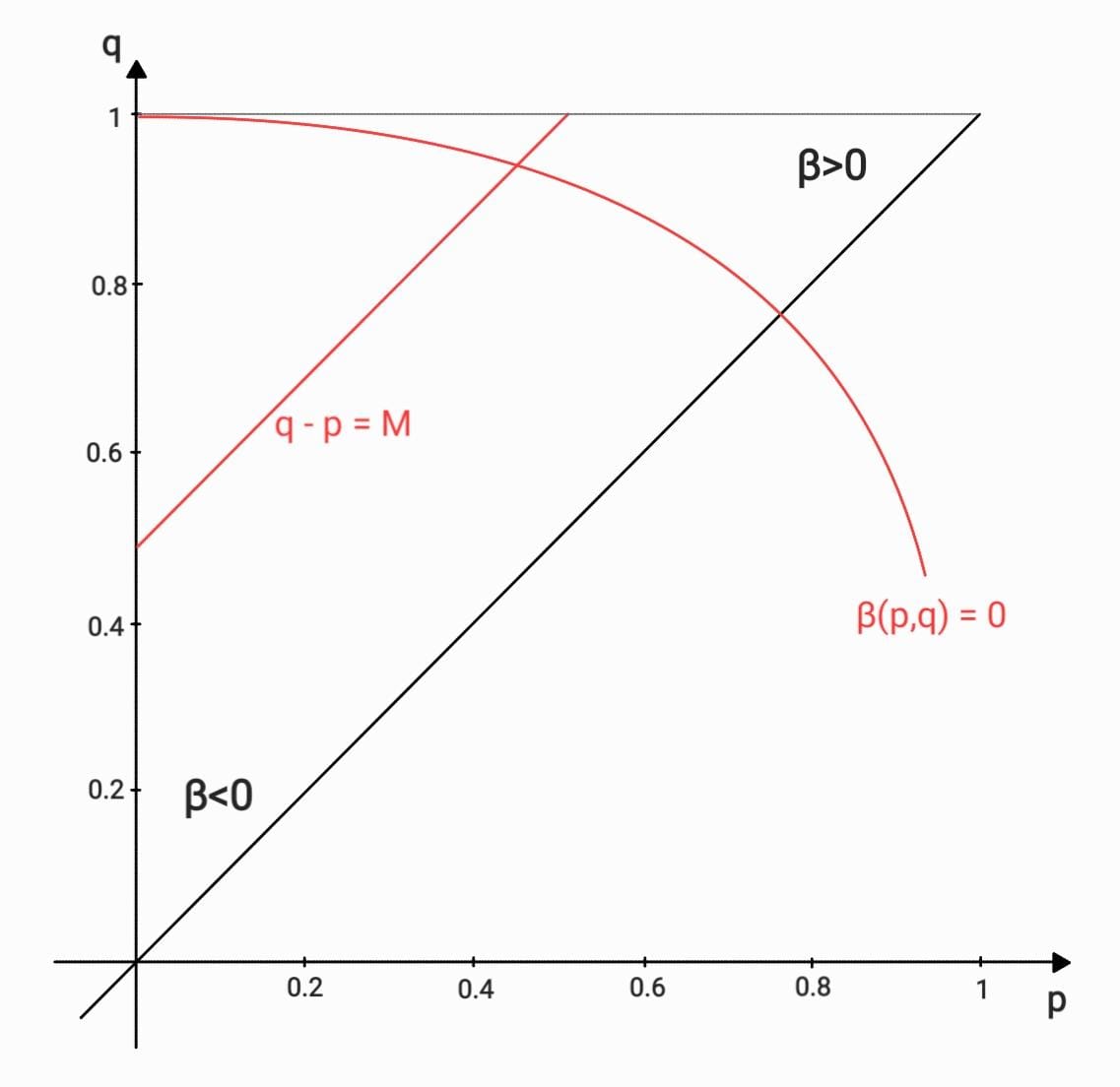}

\label{fig:theory}
\begin{flushleft}
\footnotesize 
\singlespacing
\textit{Note}: The figure displays all the intervals $(p,q)$, with $p\leq q$. The red lines are equations (\ref{eqn:def1}) and (\ref{eqn:def2}).
\end{flushleft}
\end{figure}

The intuition about the existence of a middle class is given by the extreme behavior of $\beta$, and a continuity argument. Figure (\ref{fig:theory}) displays the function $\beta(p,q)$ in the space of quantiles $(p,q)$, with $p<q$. 

\bigskip

The assumption (\ref{eqn:ineq1}) implies that $\beta$ is negative at the lower vertex of the triangle, while the assumption (\ref{eqn:ineq1}) indicates that it is positive at the upper vertex. Following continuity arguments, any path from one vertex to the other necessarily crosses a line where the condition (\ref{eqn:def1}) holds. Consequently, the line $\beta(p,q)=0$ separates the quantile space into two halves. In addition, the condition $S(0,1)=1$ implies that $\beta(0,1)=0$ and so this separating line cuts the third vertex in the upper left corner, constituting a path from $(0,1)$ to some point on the 45-degree line. In addition, the condition (\ref{eqn:ineq2}) is the locus given by the 45-degree line crossing the vertical axis in $q=M$. It is straightforward that the two lines intersect at a point.  

The following sections reproduce this figure using simulated and actual data, proving the existence of an insensitive middle class.

\subsection{The Linearity Assumption}

The previous arguments rely on the linear assumption (\ref{eqn:linear}), so all analysis is conducted on the linear coefficient $\beta(p,q)$. The assumption is testable, and in the following sections, I will show that it holds for both rich and poor classes and for theoretical and real-world distributions. However, the middle-class is defined at the point where $\beta=0$, and at this point, non-linear terms may become important. The data in the following sections also show that the linear approximation is not a good specification near the percentiles that define the middle class.

Nevertheless, the definition of the middle class can be generalized without depending on the linear assumption. The condition $\beta=0$ in (\ref{eqn:def1}) means that inequality is not affecting that specific income share. The condition $R^2=0$ is equivalent - with $R^2$ the goodness of fitting of equation (\ref{eqn:linear}) - given that for a linear univariate regression $R^2=\beta^2[\text{var}(G)/\text{var}(S)]$. Accordingly, $\beta$ closer to zero is equivalent to minimizing $R^2$, and the middle-class quantiles in which $\beta$ is closer to zero are those least explained by inequality.

The previous equivalence implies that the definition of the middle class does not require the linear assumption. The new definition replaces equation (\ref{eqn:linear}) by the general specification $S(p,q)_{i}=F(G_i; p,q)+\varepsilon_{i}$. For instance, $F(G_i; p,q)$ can be a polynomial with coefficients that depend on $(p,q)$. Denote $R^2(p,q)$ the coefficient of determination of the estimation.

\textbf{DEFINITION.} The middle class is a quantile interval $(p,q)$ of size $M$ that minimizes the proportion of the variance of income share $S(p,q)$ which is explained by the variance of $G$. Formally, it solves the following optimization problem:

\begin{subequations}
\begin{eqnarray}
& \arg\min_{\{p,q\}} \; R^2(p,q) \label{eqn:nonlinear1}  \\
& \text{s.t. } q- p = M \label{eqn:nonlinear2}
\end{eqnarray}
\end{subequations}

The expressions (\ref{eqn:nonlinear1}) and (\ref{eqn:nonlinear2}) are equivalent to conditions (\ref{eqn:def1}) and (\ref{eqn:def1}) for the linear case, and provide a general definition of the middle class. In this context, the question of whether the linearity assumption is valid is an empirical one. If the problem (\ref{eqn:nonlinear1}) has similar solutions for linear and non-linear specifications, then the linear approximation is valid. This will be the case for practical applications, as the following sections show.

\section{The Middle Class for a Pareto Distribution}

As a first exercise, I parametrize the income distribution function as a Pareto distribution. Under this parametrization, the relationship between income shares $S(p,q)$ and the Gini coefficient is nonlinear. A primary objective of the exercise is to assess whether the assumptions of the previous section, i.e., equation (\ref{eqn:linear}) and inequalities (\ref{eqn:ineq1}) and (\ref{eqn:ineq2}), hold. The section proceeds in reverse order: after discussing the Pareto parametrization, I consider the limit behavior of shares for the very poor and the very rich, showing that the previous inequalities are appropriate. Then I compute and discuss the insensitive middle class, and finally assess the validity of the linear specification for the relationship between shares and Gini. 

\subsection{Pareto Parametrization}

A society is composed of individuals with zero and positive incomes. A mass $\nu \in (0,1)$ of individuals have incomes equal to zero, while a Pareto distribution function distributes the rest. The proposed distribution is simple enough to imply a closed formula between quantiles and inequality, while capturing the basic structure of income distribution.\footnote{That the distribution of income has two parts, and the use of the Pareto distribution for the right tail, are two well-known results in this literature. See, for instance, \cite{reed2004double, toda2012double, benhabib2018skewed}.} The incorporation of individuals with zero income gives us a greater level of flexibility when modeling distributions in practice, as discussed below.

The distribution of income is given by: 

\begin{equation} 
f(y) = 
\begin{cases}
\nu & \text{if } y = 0 \\
\alpha (1-\nu) \frac{y_m^{\alpha}}{y^{\alpha +1}}  & \text{if } y > y_m
\end{cases}
\label{eqn: Pareto}
\end{equation}

A Pareto distribution with parameter $\alpha>1$ has a closed form for the Gini index, equal to $1/(2\alpha-1)$. In the case of (\ref{eqn: Pareto}), inequality should also consider the fraction $\nu$ of individuals with no income. It is easy to show that the Gini $G$ of the distribution (\ref{eqn: Pareto}) is given by $G=\nu+(1+\nu)/(2\alpha-1)$. From this expression, $\alpha$ can be written as a function of $G$ and $\nu$.

For two arbitrary percentiles $p$ and $q$, with $\nu \leq p < q \leq 1$, the income share $S(p,q)$ is given by: 

\begin{equation}
S(p,q)= \left(\frac{1-p}{1-\nu}\right) ^{\left( \frac{1-G}{1+G+2\nu} \right)} - \left(\frac{1-q}{1-\nu}\right) ^{\left( \frac{1-G}{1+G+2\nu} \right)}
\label{eqn:shares}
\end{equation}

I simulate the relationship between income shares and Gini indices, using formula (\ref{eqn:shares}). The income share $S(p,q)$ is plotted for a simulated sample of Gini indices. A normal distribution with mean $0.4$ and variance $0.05$ is an accurate approximation of the true distribution of income. The results are robust to the use of other sampling distributions. As formula (\ref{eqn:shares}) provides a one-to-one relationship between Gini and income shares, I assume that $\nu$ is also changing to add variation. The value of $\nu$ is the modal percentile, which is scarcely discussed in the literature. However, empirical data suggest that the distribution's mode is typically in the second decile. Accordingly, I assume a uniform distribution on that decile. As in the case of $G$, the election of the sampling distribution of $\nu$ is not driving the results. Therefore, the Monte Carlo exercise considers: 

\begin{eqnarray*}
G &\sim& N(0.4,0.05) \\ 
\nu &\sim& U[0.1,0.2]
\end{eqnarray*}

In what follows, we will use the convention of denoting $p$, $q$ and $M$ as percentiles, that is, between $0$ and $100$. 

\subsection{Linear Limit Behavior}

\begin{figure}[H]
    \centering
\centering
  \caption{Extreme Quantiles' Shares versus Gini Index}
  \includegraphics[width=0.8 \textwidth]{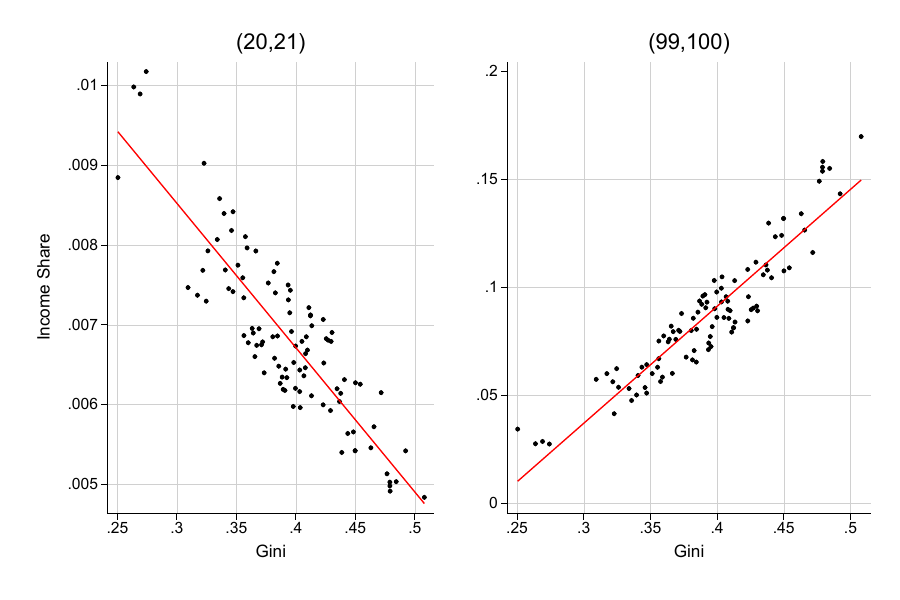}
\label{fig:shares1}
\begin{flushleft}
\footnotesize 
\singlespacing
\textit{Note}: Each point is a realization of the Monte Carlo experiment. The shares are computed by (\ref{eqn:shares}) for the respective percentiles.
\end{flushleft}
\end{figure}

First, I study the limit behavior of beta given by the assumptions (\ref{eqn:ineq1}) and (\ref{eqn:ineq2}). I consider $\varepsilon=0.01$, so the assumptions refer to the first and last percentiles in the distribution. Since the poorer members of the simulated society have zero income, I use the 20-21 percentile as the first decile. Figure (\ref{fig:shares1}) plots $(20,21)$ and $(99,100)$ against $100$ simulated Gini indices. 

As expected, the poorest (richest) percentile decreases (increases) its income share as inequality increases. Moreover, in the two cases, the linear approximation given by (\ref{eqn:linear}) is a precise approximation of (\ref{eqn:shares}). The $R^2$ is around $0.76$ for $(20,21)$, and $0.86$ for $(99,100)$. The addition of non-linear terms to the Gini does not improve the goodness of fit. 

\subsection{The Middle Class}

\begin{figure}[H]
    \centering
\centering
\caption{Regression coefficients for Quantiles}
  \includegraphics[width=0.8 \textwidth]{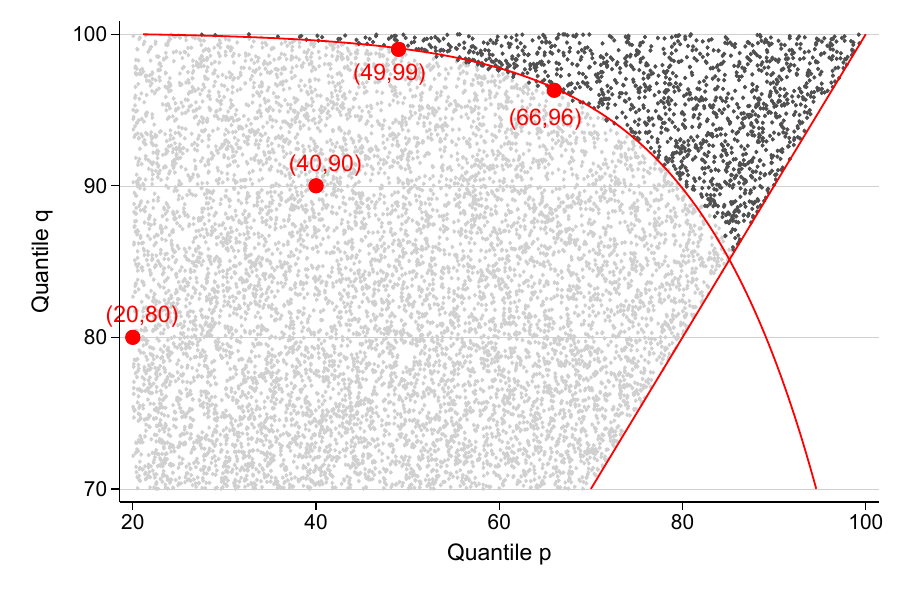}
\label{fig:allshares}
\begin{flushleft}
\footnotesize 
\singlespacing
\textit{Note}: The figure display in black (gray) the positive (negative) $\beta$ coefficient from the estimation of equation (\ref{eqn:linear}). The quantiles $(p,q)$ are $10,000$ random draws using the same $100$ sample points of the Monte Carlo experiment. 
\end{flushleft}
\end{figure}

The simulation computes the shares for arbitrary pairs of quantiles $(p,q)$. For each of them, I estimate the beta of the regression (\ref{eqn:linear}). That maps the beta coefficient in the entire space of quantiles, as in Figure (\ref{fig:theory}). The black points are those quantiles $(p,q)$ with $\beta>0$; gray points, in contrast, indicate $\beta<0$. Figure (\ref{fig:allshares}) displays the results.

The figure exhibits the expected theoretical pattern. The estimated beta is negative for low quantile pairs, the poorer segment of the income distribution. In contrast, the estimated betas are positive for the higher ones, the richer groups. In the middle, there is a frontier line where $\beta=0$. 

\begin{figure}[H]
    \centering
\centering
\caption{Middle Class Shares versus Gini Index}
  \includegraphics[width=0.8 \textwidth]{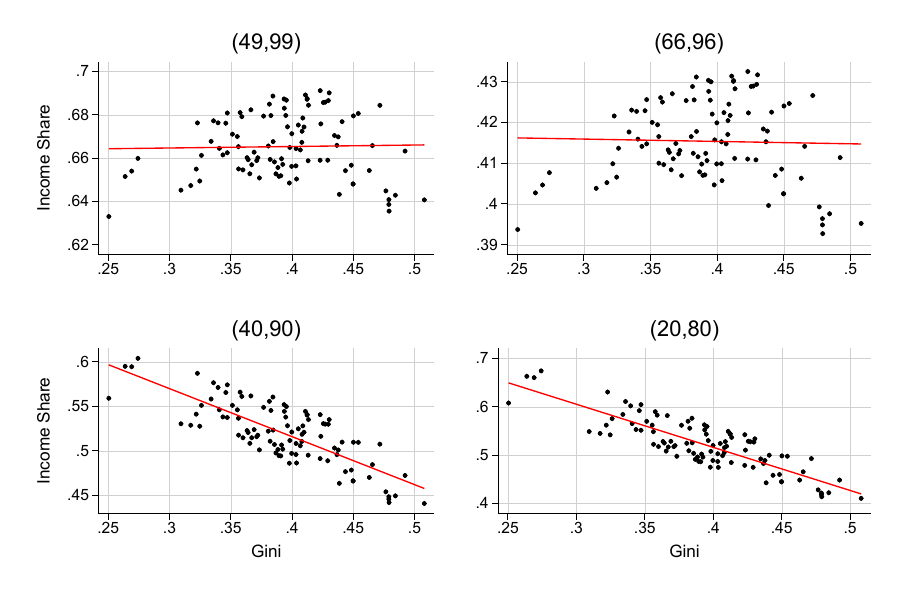}
\label{fig:shares2}
\begin{flushleft}
\footnotesize 
\singlespacing
\textit{Note}: Each point is a realization of the Monte Carlo experiment. The shares are computed by (\ref{eqn:shares}) for all the middle classes. 
\end{flushleft}
\end{figure}

The red line represents the middle class; the value of M single out a particular one. For $M=50$, the quantiles $(49,99)$ define the middle class; for $M=30$, the middle class is $(66,96)$. Notably, other middle classes in the literature are on the left side of the red frontier, indicating a negative relationship with Gini. For comparison, the Easterly middle class $(20,80)$ and the Palma middle class $(40,90)$ are also plotted against inequality in Figure (\ref{fig:shares2}). 

The two middle classes in Figure (\ref{fig:shares2}) show regression slopes close to zero, with t-values well below 1. In contrast, the slopes are unambiguously negative for the standard middle classes, at the bottom of the figure. The correlation of the group $(20,80)$ shares with Gini is $-0.85$, while the correlation of the shares of the Palma middle class, $(40,90)$, is $-0.79$. \cite{cobham2016inequality} reported a similar correlation between the share of the Palma middle class and inequality. The coefficient of variation is also lower for the middle classes of the red line. Comparing the $M=50$ middle classes, $(49,99)$ has the lowest coefficient of variation of all other classes of the same size, with $CV=0.021$, while $(40,90)$ has $CV=0.065$, three times higher.

\subsection{The Linearity Assumption}

Finally, I discuss the validity of the linear specification (\ref{eqn:linear}) in the context of a Pareto distribution. As a null linear term defines the middle class, the emergence of non-linear terms is expected in those percentiles. In fact, Figures (\ref{fig:shares2}) suggest a mild quadratic relationship which is not incorporated in equation (\ref{eqn:linear}). As a matter of fact, the linear assumption can be easily disputed by the following argument. In the case of $G=0$, $\alpha(p,q)=q-p$ because of perfect equality. For the middle class, this implies the mean share equals $M$. However, Figure (\ref{fig:shares2}) shows that mean shares are close to $0.66$ and $0.42$ for $M=50$ and $M=30$, respectively. The discrepancy suggests that the relationship (\ref{eqn:linear}) is non-linear near the percentiles that define the middle classes. 

\begin{figure}[H]
    \centering
\centering
\caption{R-squared for Linear, Quadratic and Polynomial Specification}
  \includegraphics[width=0.8 \textwidth]{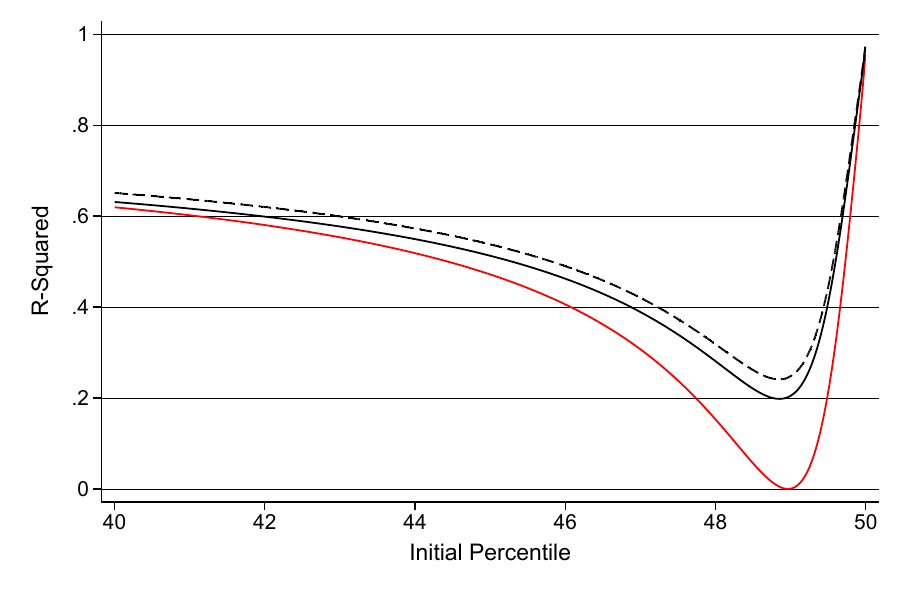}
\label{fig:figR2}
\begin{flushleft}
\footnotesize 
\singlespacing
\textit{Note}: R-squared for several specifications, for the interval $(p,p+50)$ with p the initial percentage. The red line is the linear specification, the black line is quadratic, and the black dashed line is the specification with a fifth-order polynomial in Gini.
\end{flushleft}
\end{figure}

However, as explained in the previous section, the linear specification can be generalized using non-linear terms. In such a case, the definition of the middle class relies on the minimization of the statistic $R^2(p,q)$. For simplicity, I focus on $M=50$. I consider the $R^2$ of the linear regression (\ref{eqn:linear}) for intervals whose star is between $40$ and $50$, including the previously computed endogenous middle class at $49$. In addition, I plot the $R^2$ of two other estimations: the first includes a quadratic term, and the second is a polynomial of order 5 in Gini. The Figure (\ref{fig:figR2}) displays the results.

First, we observe that for the linear case, $R^2$ is minimum close to 49, as expected. Moreover, the value is (almost) zero, indicating the beta is (almost) zero, too. Secondly, the incorporation of Gini squared increased $R^2$, particularly near the minimum, supporting the intuition that non-linearity matters when the linear term vanishes. Third, the addition of extra terms, such as cubic terms, increases the goodness of fit only slightly compared to the quadratic one.  

The key point of the figure, however, is that the middle class implied by $\beta$ close to zero, or equivalently $R^2$ close to zero in the linear equation, is almost the same as the one implied by non-linear equations. In the first case, the exact left side of the middle class is the percentile $48.96$. In the two other cases, the minimum is located at $48.85$. In practice, these two values are indistinguishable. The conclusion is that the linear equation (\ref{eqn:linear}) is valid to compute the middle class unresponsive to inequality. 

\section{The World Middle Class: 1980-2024}

This section computes the middle class using data from the World Inequality Database (WID). WID has detailed income share data for all percentiles across 215 countries since 1980, encompassing almost a million data points. I use the complete sample of country-year observations to compute the middle class. Next, I consider other samples: cross-sectional samples across different years and country-specific samples. The results are very consistent across both the between- and within-country samples. 

\subsection{Word Inequality Database (WID)}

I will compute the middle class worldwide, using information from the World Inequality Database (WID). In recent years, WID has become a cornerstone resource for analyzing trends in income and wealth inequality \citep{alvaredo2017global, chancel2022world}. Created by researchers who began studying top incomes and are associated with the World Inequality Lab, the WID provides transparent, high-quality, and internationally comparable inequality data. Unlike many traditional data sources that rely solely on household surveys, WID combines tax records, national accounts, and surveys to provide more consistent estimates of inequality. The approach enables capturing high-income individuals, who are often underrepresented in surveys, making it particularly useful for studying top-end income and wealth concentration. 

Specifically, WID provides detailed disaggregation of income and wealth across the distribution, with estimates available at the percentile level. This granularity enables a detailed analysis of the income distribution, as presented in this paper. The data used are percentile distributions of post-tax income, which measure individual disposable income. 

The sample includes 215 countries worldwide, with coverage from 1980 to 2023. The total number of country-year units is $9,460$. The calculations for the middle class use the entire sample, but some figures do not include outliers.\footnote{The three countries excluded in some figures are Malawi, Maldives, and Yemen. These countries present data for some income shares that are more than five standard deviations from the country averages. When excluded, the reported number of observations is $N=9,328$. All the results, in any case, are robust to their inclusion/exclusion.} 

\subsection{Computation of the Middle Class}

This section considers the entire sample of country-year units to construct the middle class. For each of these units, WID reports the 100th percentile; the total number of observations used in the analysis is close to 1 million.

To check the inequalities (\ref{eqn:ineq1}) and (\ref{eqn:ineq2}), I regress the Gini index on the income shares of the 1st and 100th percentiles. For the first percentile, the beta coefficient (std.error) is $-0.00688 (0.0007)$ and $R^2=0.525$. For percentile 100, $\beta(99,100)= 0.3990 (0.0020)$ and $R^2=0.804$. 

\begin{figure}[H]
    \centering
\centering
\caption{The Middle Class}
  \includegraphics[width=0.8 \textwidth]{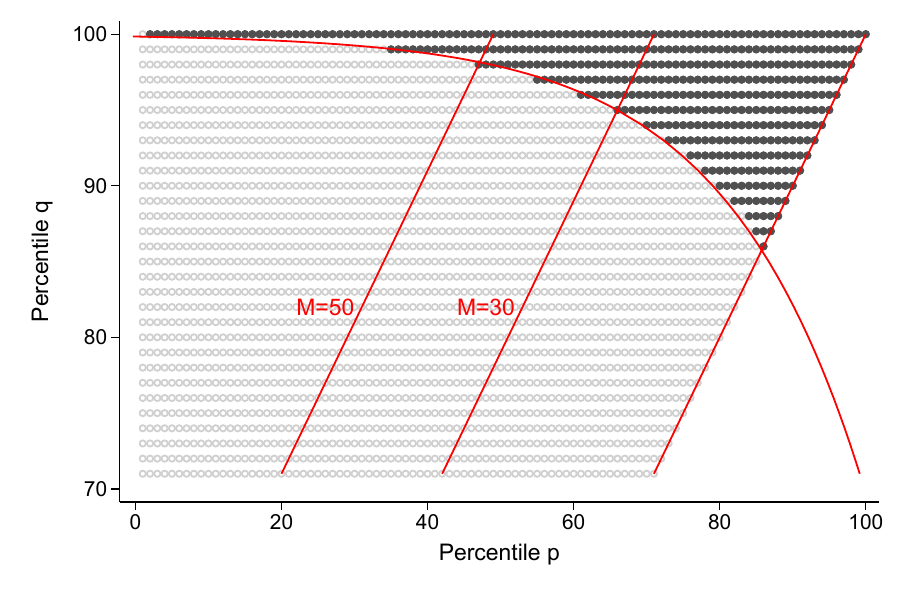}
\label{fig:allshares_wid}
\begin{flushleft}
\footnotesize 
\singlespacing
\textit{Note}: The figure displays in black (gray) the positive (negative) $\beta$ coefficient from the estimation of equation (\ref{eqn:linear}) using WID data. For each point, the regression includes $N=9,460$ country-year units.   
\end{flushleft}
\end{figure}

Figure (\ref{fig:allshares_wid}) reproduces the coefficient of regressing Gini on income share for each pair $(p,q)$ of percentiles. 

In the Figure, dark points indicate a positive and significant beta, while the rest are negative. The pattern of actual data is similar to figures (\ref{fig:theory}) and (\ref{fig:allshares}). The lower percentiles decrease their income shares as inequality increases, while the opposite occurs among the richer percentiles. There is a zone of percentiles where $\beta(p,q)$ is close to zero. Imposing a specific size, the equations (\ref{eqn:def1}) and  (\ref{eqn:def2}) define a unique set of percentiles as the engogenous middle class.

To focus the analysis, the Figure displays two middle classes of different sizes. The first one is the $M=50$ middle class, given by $(48,98)$.\footnote{In terms of percentiles, the middle class begins at $49$ because that percentile is defined as the interval $]48,49]$.} Secondly, the $M=30$ middle class is given by $(65,95)$. While these classes are directly identifiable by inspection, they can be formally calculated by minimizing $\beta^2$ or $R^2$ in the regression (\ref{eqn:linear}), for a fixed M. 

\begin{figure}[H]
    \centering
\centering
  \caption{Poor, Middle and Rich Classes vs. Gini}
  \includegraphics[width=0.8 \textwidth]{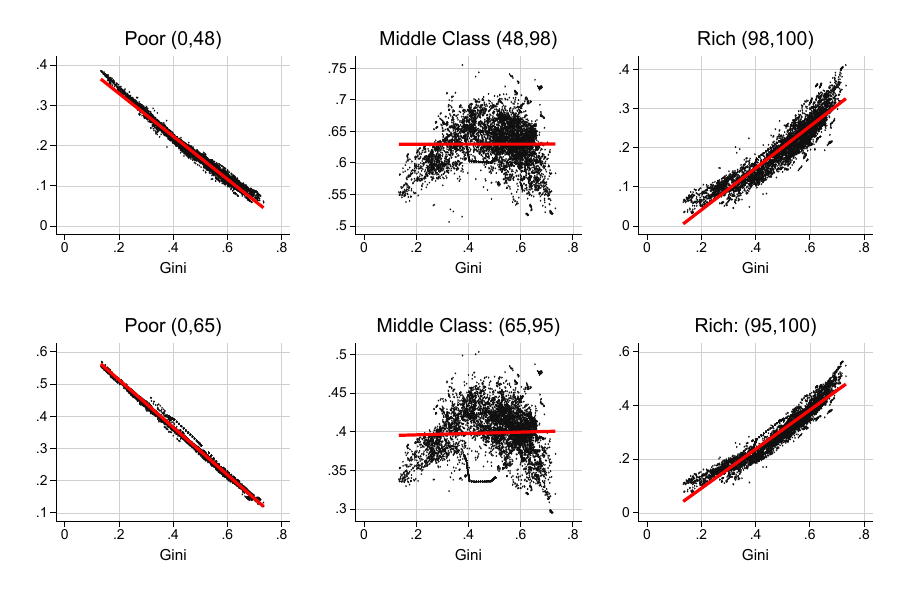}
\label{fig:classes_gini}
\begin{flushleft}
\footnotesize 
\singlespacing
\textit{Note}: Observations are $N=9,328$ country-year units between 1980 and 2023. 
\end{flushleft}
\end{figure}

The middle-class intervals define the other two classes, the poor and the rich. For $M=50$ and $M=30$ thresholds, Figure (\ref{fig:classes_gini}) displays the relationship of Gini with all classes in the distribution.

The Figure shows that the three classes are characterized by their correlation with inequality. The poor and the rich's income shares strongly correlate with the Gini, though in opposite directions. The middle class, in contrast, exhibits no linear relationship.\footnote{As discussed in the previous section, the middle class also depends non-linearly on Gini. However, the minimization of $R^2$ for a general polynomial function of $G$ results in the same middle classes discussed here.} 

The figure shows that while the middle class does not vary with the Gini coefficient, it does exhibit variance. Admittedly, the middle-class variance is lower than that of the other classes. For $M=50$, for instance, the coefficient of variation of the poor and rich classes is $40.5$ and $34.5$, respectively, while dropping to $5.1$ for the middle class. Yet contrary to \cite{palma2006globalizing, palma2011homogeneous, palma2019behind}, the middle classes are not fixed but differ in their income shares. 

The standard middle classes exhibit results consistent with those discussed in the previous section. The $(20,80)$ middle class correlates  $-0.95$ with inequality, while the Palma middle class, $(40,90)$, correlates $-0.87$ with Gini. As for the coefficient of variation, those classes exhibit greater variance. In particular, the Palma invariant middle class has a coefficient of variation of $13.1$ in all the sample, more than twice that of the $(48,98)$ middle class. 

\subsection{The Size M}

\begin{figure}[H]
    \centering
\centering
  \caption{M-Size Middle Classes}
  \includegraphics[width=0.8 \textwidth]{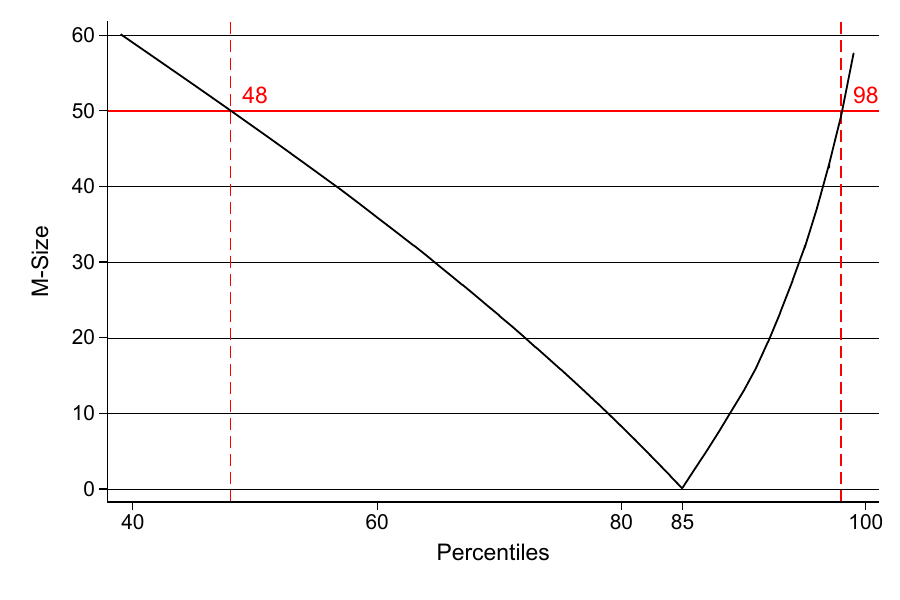}
\label{fig:parabola}
\begin{flushleft}
\footnotesize 
\singlespacing
\textit{Note}: The downward-sloping (upward-sloping) line is the initial (final) percentile of the middle class for any size $M$. The $M=50$ middle class is in the red line.
\end{flushleft}
\end{figure}

Figure (\ref{fig:allshares_wid}) exhibits the middle class frontier, given by $\beta(p,q)=0$, and illustrates two middle classes for $M=30$ and $M=50$. The middle class size is arbitrary, though high levels are problematic given that the middle class is close to the right-handright-hand side of the distribution. The Figure (\ref{fig:parabola}) displays the initial and final percentiles of all middle classes from $M=1$ to $M=60$.

In the Figure, the $M=50$ middle class $(48,98)$ is shown in red. However, any M-size middle class can be singled out. The percentile $85$ plays an important role, as it falls within the middle class of every M-size. For $M=1$, that is the middle class, namely the percentile with the lower response to inequality. The Figure is asymmetric, adding more low percentiles for each increase in $M$. 

\begin{figure}[H]
    \centering
\centering
  \caption{M=50 and M=30 Middle Classes}
  \includegraphics[width=0.8 \textwidth]{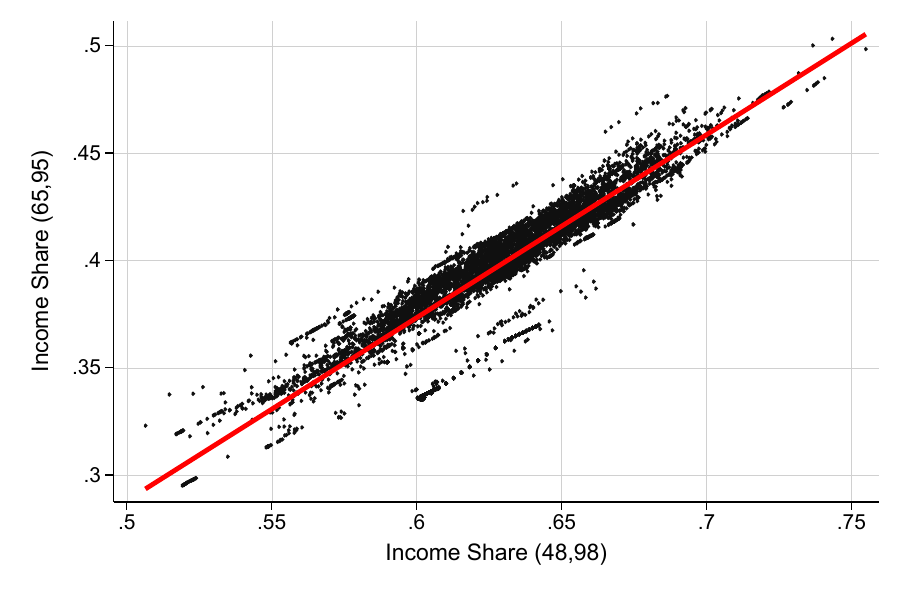}
\label{fig:correlation}
\begin{flushleft}
\footnotesize 
\singlespacing
\textit{Note}: Points are all countries between 1980 and 2023. $N=9,328$. 
\end{flushleft}
\end{figure}

The arbitrariness of $M$ is an obvious concern, given that the middle class, as measured by its share, depends on the choice of $M$. To explore differences among the M middle classes, Figure (\ref{fig:correlation}) displays the relationship between income shares for $M=50$ and $M=30$ across the entire sample of countries from 1980 to 2013. 

The correlation between the two middle classes is $0.90$. The correlation is also high for the other M-size middle classes. For instance, M=10 correlates $0.77$ and $0.96$ with $M=50$ and $M=30$, respectively. The result is reassuring, given the arbitrariness of $M$. However, whether different M provide different statistical results is a question that the specific econometric analysis of certain problems will have to answer.

\subsection{Cross-country Middle Classes}

The definition of the middle class is sample-dependent. The middle classes discussed so far were calculated using the entire sample of countries, for all available years (1980-2023). The natural question is how those values compare to other samples. First, we will consider the middle class for a fixed year across a cross-section of countries. This between-country calculation allows examination of the variance in the definition over time. In addition, the middle class definition depends on the measure of inequality used. A robust definition of the middle class should be invariant to the measure used.

\begin{figure}[H]
    \centering
\centering
  \caption{Beta Coefficients for Cross-Country Samples}
  \includegraphics[width=0.8 \textwidth]{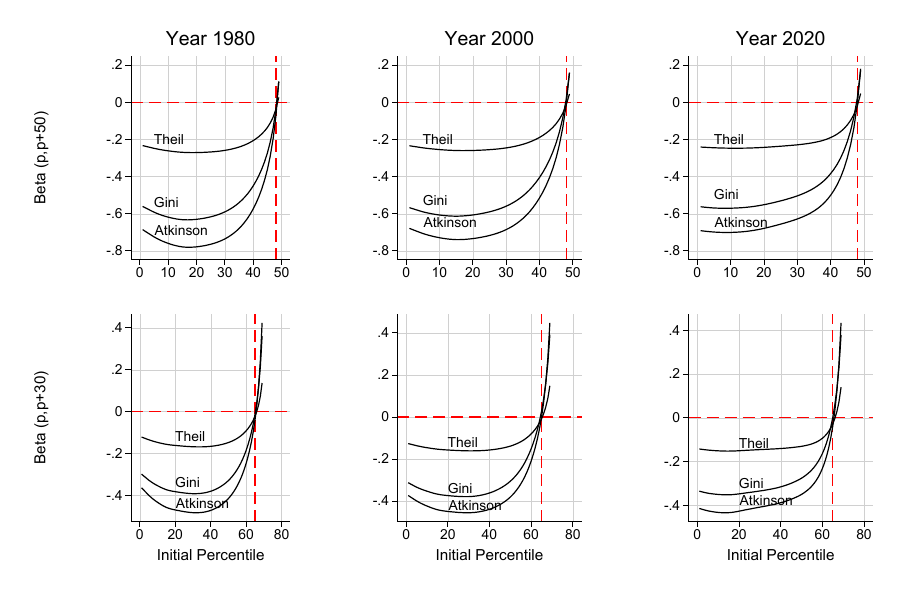}
\label{fig:cross_section}
\begin{flushleft}
\footnotesize 
\singlespacing
\textit{Note}: Beta Coefficient as a function of the initial percentile of the interval, in three moments of time. $M=50$ above; $M=30$ below. 
\end{flushleft}
\end{figure}

Therefore, I compute cross-country middle-class shares over time for several measures of inequality. Figure (\ref{fig:cross_section}) displays the results. I consider the entire period with data, from 1980 to 2020, every 20 years, and three measures of inequality: Gini, Atkinson, and Theil. The figure shows the estimated beta for $M=50$ and $M=30$ classes, for each year and inequality measure.

The Figure exhibits remarkable robustness over time and inequality measures. For $M=50$, at the top of the Figure, the three years exhibit the same pattern, with beta crossing zero at the $(48,98)$ middle class. For $M=30$, the Figure displays the same pattern.  

The stability of the middle class is reassuring. When commenting on the Palma index, Milanović correctly points out that "what is immobile may change between the countries, or across time". In this sense, the coverage of the WID database is fundamental for showing the stability of the middle class since 1980, a period in which inequality has increased sharply. Whether the middle class, as defined endogenously in this work, has been historically stable is an important question that requires further information to be answered.

\subsection{Country-specific Middle Classes}

Secondly, I compute the country-specific middle classes. The literature has considered definitions of the middle class that vary between countries. For instance, researchers agree that the notion of the middle used in developed countries is not trivially exported to developing ones \citep{milanovic2002decomposing, banerjee2008middle, ravallion2010developing, lopez2014vulnerability, rasch2017measuring}. Here, we will go a step further and define a middle class for each country.   

\begin{figure}[H]
    \centering
\centering
  \caption{Histogram Country-specific Middle Classes}
  \includegraphics[width=0.8 \textwidth]{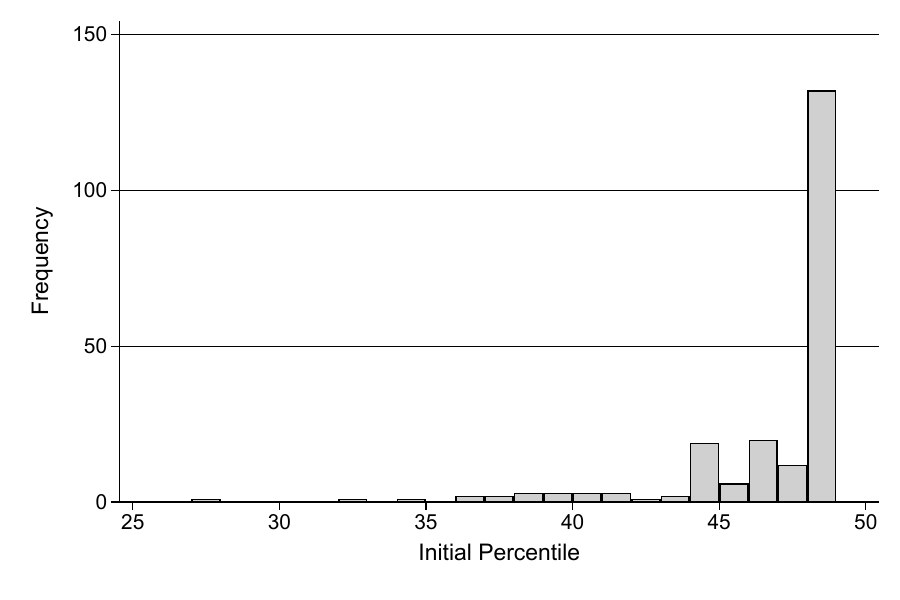}
\label{fig:hist_perc}
\begin{flushleft}
\footnotesize 
\singlespacing
\textit{Note}: Initial percentile of the county specific middle class. 211 countries. 
\end{flushleft}
\end{figure}

To compute country-specific middle classes, the estimation in (\ref{eqn:linear}) uses samples from 1980 to 2023 for each country, with $N=44$. The middle class is the percentile interval with size $M=50$ that minimizes $R^2$. In a few cases, the optimization results in an extreme interval, so I restrict the initial $p$ to be strictly greater than zero and strictly less than $50$ to include both poor and rich classes. The Figure (\ref{fig:hist_perc}) summarizes the initial percentile of these middle classes\footnote{In the histogram, Mexico is excluded because it is an outlier: (1,51). However, when imposing $M=49$ or $M=51$, the initial percentile of the Mexican middle class is $49$.}

Country-specific middle classes are skewed to the right of the distribution, as is the global middle class. A significant number of countries have middle classes in the interval (49-99). A large fraction of the rest initiated the interval above 40. The mean initial percentile is $46.8$. While country-specific middle classes exhibit some variance, the previous figures do not deviate too much from the global middle class $(48.98)$.

Finally, the other inequality measures show a similar pattern, although the numbers are not identical. For Atkinson and Theil, the mean initial percentiles of the country-specific middle classes are $46.9$ and $44.2$, respectively. Also, for those measures, a large number of countries have the middle class in the interval $(49,99)$. While in the Gini case described by Figure (\ref{fig:hist_perc}), the fraction of those countries is $53.0\%$, the percentages are $56.9$ and $55.4$ for Atkinson and Theil, respectively.

\section{The World Middle Class: a description}

This section describes the middle-class inequality insensitive computed in the previous section. First, such a middle class, for any M-size, is skewed toward the rich. Secondly, it exhibits variance in income shares. Thirdly, I discuss the evolution of the endogenous middle class in comparison with the other standard middle classes. 

\subsection{Percentiles}

The immediate observation about the insensitive middle class is that it is not close to the median income of the distribution but skewed towards the rich percentiles. It includes all the richer segments, except for the very rich. These results are for all M-size, as Figure (\ref{fig:parabola}) illustrates. The $M=10$ middle class is practically the ninth decile. For higher $M$, the middle classes include significant fractions of the last decile. About the lower bound, for $M=50$, the lower bound of the middle class is slightly below the median. For smaller sizes, that group is strictly to the right of the median. 

A middle class skewed to the rich contrasts with part of the existing literature. As explained, the standard criterion for setting the lower threshold of the middle class is the poverty line. Absolute income thresholds are just above the survival level, especially in studies of developing countries. In the case of the relative income middle class, the preferred approach considers the middle class as the interval between $75\%$ and $200\%$ of median income; that implies average percentiles of $35$ and $90$ in the WID 2023 sample. In addition, the standard middle classes hardly include the last decile. All those middle classes are to the left of the ones discussed here.  

However, other literature suggests that the middle class is related to affluent citizens, except for the very rich. Although he favors a definition based on wealth, Piketty designated the middle class as a $(50,90)$ interval. While he acknowledge that this designation is obviously arbitrary and open to change, he explains that those figures correspond more closely to the common usage of the term: "the expression middle class is generally used to refer to people who are doing distinctly better than the bulk of the population yet still a long way from the true elite" \cite[p.251]{piketty2014capital}. Also, the invariant Palma middle class bias towards the rich \citep{palma2011homogeneous}. \cite{birdsall2010indispensable} defines the upper threshold of the middle class as the 95th percentile; her middle class is uncorrelated with inequality. Finally, \cite{milanovic2002decomposing} described the world's middle class by noting that $78$ percent of the world's population is poor, $11$ percent belongs to the middle class, and $11$ percent are rich. Remarkably, these cutoffs are precisely the ones implied by the election $M=11$, where the middle class interval is $(78,89)$.  

A relatively wealthy middle class is closer to this group's self-perception. Surveys show that self-adjudicated middle classes are to the right of the median. Typically, the percentage of individuals who report themselves as rich or upper classes is between $1\%$ and $3\%$ \citep{eisenhauer2008economic, cashell2008middle}. At the time, households are within the middle class if their incomes fall between the 45th and 99th income percentiles \citep{pressman2007decline}.\footnote{The report Americans' social Class Self-Identification, by Gallup, reports similar results for 2002-2024. Only $2\%$ of US citizens listed themselves as upper class. The Pew Research Center, in contrast, reports on a much larger upper class.}


\subsection{Income Shares}

\begin{figure}[H]
    \centering
\centering
  \caption{Middle Class Income Shares}
  \includegraphics[width=0.8 \textwidth]{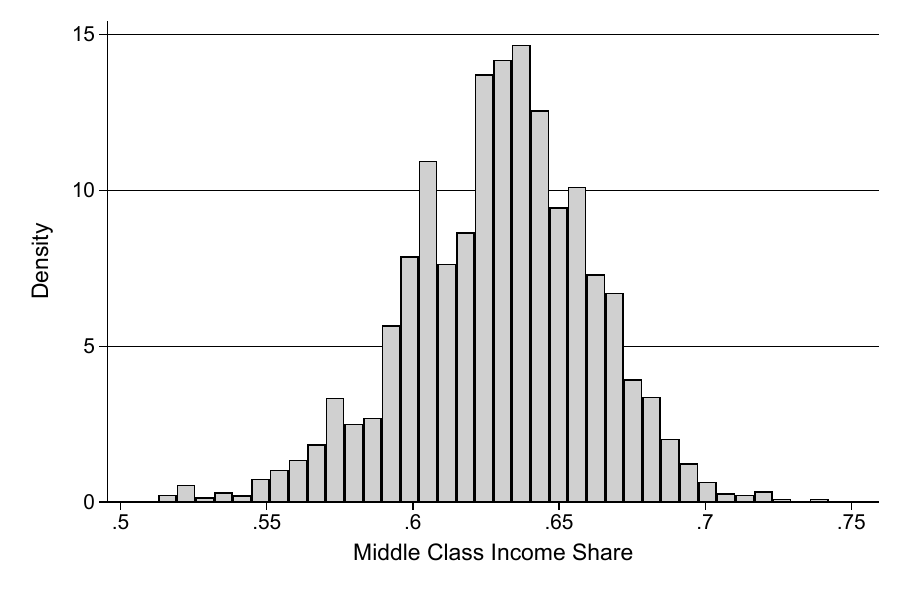}
\label{fig:fig_MC}
\begin{flushleft}
\footnotesize 
\singlespacing
\textit{Note}: Middle Class is (48,98). The unit is country-year. $N=9,328$. 
\end{flushleft}
\end{figure}

The definition of the middle class based on fixed percentiles implies that the middle-class variance is determined by the income shares held by those percentiles. While the entire sample of country-year units was used to compute the percentiles, once fixed, each unit has a unique middle class's income share. The Figure (\ref{fig:fig_MC}) displays the histogram of countries' income shares for the period 1980-2023.

The histogram is relatively symmetric, although the null hypothesis of skewness is not rejected. The mean is $0.631$, and the median is $0.629$. The range of observations is $[0.506,0.755]$, with 8 out of 10 countries having middle-class shares between $0.590$ and $0.666$, and half within a five percent window.\footnote{However, outliers have been dropped. Malawi in the 80s had figures close to $0.30$.} 

For $M=30$, the middle class $(65,95)$ has a mean income share of $0.400$, with half of the countries between $0.373$ and $0.416$.

\subsection{Evolution of the middle class}

Since 1980, inequality has evolved in an inverted U shape \citep{chancel2022world}. In our data, the average Gini increased from 0.508 to 0.521 in 2007, and then decreased below 0.5 at the end of the period. 

\begin{figure}[H]
    \centering
\centering
  \caption{Evolution of the Standard Classes}
  \includegraphics[width=1.0 \textwidth]{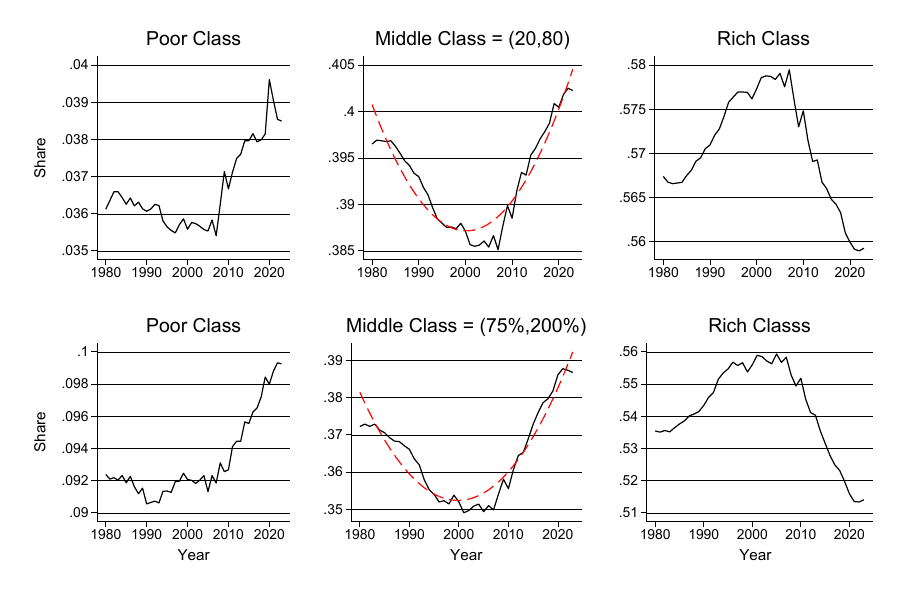}
\label{fig:evolutionSMC}
\begin{flushleft}
\footnotesize 
\singlespacing
\textit{Note}: Income shares of the poor, middle and rich classes. Yearly averages across 215 countries. 
\end{flushleft}
\end{figure}

Given their strong correlation with inequality, standard middle classes capture a dynamic similar to that of the Gini coefficient. Figure (\ref{fig:evolutionSMC}) displays the evolution of the (20,80) middle class, and the middle class that ranges from $75\%$ of the median income to $200\%$ of the median. In addition, the Figure displays the evolution of the poor and rich classes.

As announced, the two standard middle classes evolved as a U-shape, given their strong correlation with inequality. The poor follow a similar pattern, while the opposite occurred among members of the wealthy class. Interestingly, there is no global shrinking of the middle class. For 215 countries, only 99 decrease their $(20,80)$ income share in the period 1980-2023, and 92 their $(75\%, 200\%)$ share. Those countries are mainly the developed ones. 

\begin{figure}[H]
    \centering
\centering
  \caption{Evolution of the Endogenous Classes}
  \includegraphics[width=1.0 \textwidth]{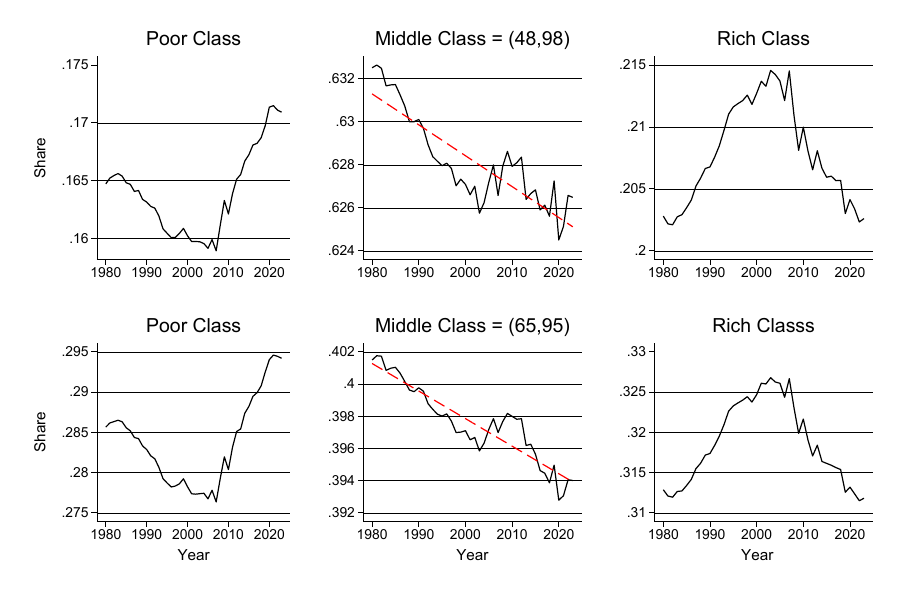}
\label{fig:evolutionIMC}
\begin{flushleft}
\footnotesize 
\singlespacing
\textit{Note}: Income shares of the poor, middle and rich classes. Yearly averages across 215 countries. 
\end{flushleft}
\end{figure}

Figure (\ref{fig:evolutionIMC}) shows a similar evolution for the endogenous middle and the other two extreme classes. 

In the Figure, the poor and rich classes show the opposite evolution over time. In contrast, the correlation in the growth rate between the middle and the poor and the rich classes is null. The two middle classes in the Figure have steadily decreased over time. For $(48,98)$ middle class, 140 out of 215 countries, or $65\%$ of the total countries, had a smaller income share in 2023 than in 1980. The fraction is reduced to $60\%$ for the $(65,95)$ middle class.

\section{Middle Class and Democracy}

Having defined the middle class, this final section illustrates how empirical studies might benefit from exploiting the information contained in the proposed measure. Since a middle class insensitive to inequality aligns with the political-economy idea of moderation, a first step is to examine the association between the size of the middle class and democracy. The goal of these estimates is not to \textit{validate} the specific conceptualization of the middle class given here, but rather to demonstrate its practical empirical implications.

There is a large body of literature examining the effect of the middle class on democracy. The idea that a strong middle class promotes democratic development dates back to Lipset’s modernization theory and Moore’s class-coalitional framework \citep{lipset1959some, moore1966social}. In contrast, \cite{huber1993impact} argue that the middle class holds an ambivalent position toward democratization. According to \cite{boix2003democracy}, the middle class matters primarily because it moderates inequality, thereby fostering democratic stability. \cite{acemoglu2006economic} develop formal models linking middle-class expansion to democratic transitions, but also highlight that the middle class can play an ambiguous and context-dependent role in these processes.

The empirical relationship between income distribution—including the size of the middle class—and democracy is inherently complex. Distributional patterns may affect democratic outcomes through various mechanisms, yet democratic institutions themselves are likely to influence income distribution. The analysis that follows does not aim to provide causal identification, but rather to establish empirical associations between variables using standard approaches widely applied in the literature.

\subsection{Data and Specification}

Both the definition and measurement of democracy are contested. Whereas \cite{dahl1971polyarchy} conceptualizes democracy as a continuous mix of participation and contestation, \cite{przeworski2000democracy} advocates a minimal, dichotomous view in which democracy is a system where parties can lose elections. There is also substantial disagreement regarding how democracy should be measured (see \cite{munck2002conceptualizing, coppedge2008two, boese2019not} for detailed discussions).

To overcome the lack of consensus in prevailing definitions of democracy, this paper relies on several alternative measures. The first is the well-known Electoral Democracy Index from the Varieties of Democracy (V-Dem) project \citep{coppedge2011conceptualizing, coppedge2019methodology}. This index provides a continuous measure, ranging from 0 to 1, of a country-year’s adherence to the ideal of electoral democracy. Conceptually, it captures the extent to which political leaders are chosen through elections conducted under broad suffrage, free and fair electoral procedures, and robust guarantees of freedom of association and expression. The index is constructed from multiple indicators corresponding to Dahl’s notion of polyarchy, based primarily on assessments by country experts and are combined using a Bayesian item-response model that adjusts for coder-specific bias and places all units on a comparable scale.

Despite its widespread use, the V-Dem Index has also been subject to criticism, particularly regarding potential coder bias, limits to cross-national comparability, and its implicitly normative conception of democracy. I therefore complement it with alternative measures. The first is the POLITY score - Polity 5 in its most recent version \citep{marshall2020polity5} - which measures the difference between the Polity Democracy and Autocracy indices. This measure ranges from $-10$ to $10$ and captures the extent of constraints on executive authority using five institutional indicators. Secondly, I employ the dichotomous Democracy–Dictatorship regime variable, introduced by \cite{alvarez1996classifying} and \cite{cheibub2010democracy}, and subsequently updated and extended by \cite{bjornskov2020regime}. This measure classifies regimes strictly as democracies or dictatorships based on institutional criteria such as contested elections and alternation in power.

Regarding the specification, an influential contribution to the estimation of large-$N$ country panels for studying the determinants of democracy is \cite{barro1999determinants}. Barro estimates his models using seemingly unrelated regressions, implicitly allowing for correlation in the country-level error terms over time. However, the standard empirical approach to analyzing democracy dynamics in country–year panel datasets is \cite{acemoglu2008income}. In their evaluation of modernization theory -namely, the effect of income on democracy - they argue convincingly that the specification must include country fixed effects. These fixed effects serve as proxies for long-run institutional features that, while constant over time, are likely correlated with both income and democracy.

Following this approach, the specification relies on five-year intervals to mitigate autocorrelation concerns and includes lagged democracy, as well as time- and country-fixed effects. I consider the following equation:

\begin{equation}
\text{DEM}_{it}=\rho \text{DEM}_{it-1}+\beta \text{MIDCLASS}_{it-1}+\gamma X_{it-1}+\delta_t+\delta_i+\varepsilon_{it} 
\label{eq:specification}
\end{equation}

where $\text{DEM}_{it}$ is an index of democracy in country $i$ at time $t$, MIDCLASS is a measure of the middle class, $X$ are controls, $\delta_t$ and $\delta_i$ represent time and country fixed effects respectively, and $\varepsilon_{it}$ is a disturbance term, assumed to be clustered by country.

The primary question concerns the statistical significance of the coefficient $\beta$ across different measures of democracy and the middle class. As noted above, there are strong reasons to condition on cross-country differences, given that country-specific factors, such as institutions, culture, or religion, may affect both democracy and the middle-class income share. For the linear specification (\ref{eq:specification}), the fixed-effect estimator identifies coefficients from within-country variation over time. However, the linear specification is not appropriate in the presence of censored or binary democracy measures \citep{benhabib2013income}. As V-DEM is a continuous, uncensored measure, I estimate (\ref{eq:specification}) directly using within-country variation. While I report linear estimates for the alternative democracy measures for comparability, the results were also verified using suitable non-linear estimators.\footnote{Available upon request.}

\subsection{Income Percentiles and Democracy}

We begin the study of the relationship between income distribution and democracy from an agnostic perspective. Rather than grouping the distribution into classes, we first examine the association between each percentile's income share and the different measures of democracy. In other words, we look at the relationship point by point along the distribution. The granularity of the WID data enables this data-driven approach, allowing us to identify general patterns that we will later formalize once we move to class-based measures.

I estimate equation (\ref{eq:specification}) replacing MIDCLASS with the percentile-specific income share. The right-hand-side variable is therefore the income share of percentile $q$ in country $i$ at time $t-1$. For each measure of democracy, I estimate the coefficients for percentiles $q = 1, \ldots, 100$. To ensure comparability, the income shares for each percentile are standardized within the sample - mean zero and standard deviation one. As for controls, I consider

As controls, I include GDP per capita (in logs; source: World Bank), a key determinant of democracy according to modernization theory. I also include inequality—measured by the Gini coefficient and its squared term—given its central role in shaping the size and composition of the middle class. Finally, I control for average years of schooling (source: Barro and Lee dataset), which strongly covaries with both middle-class size and democratic outcomes.

\begin{figure}[H]
    \centering
\centering
  \caption{Percentiles and Democracy}
  \includegraphics[width=1.05 \textwidth]{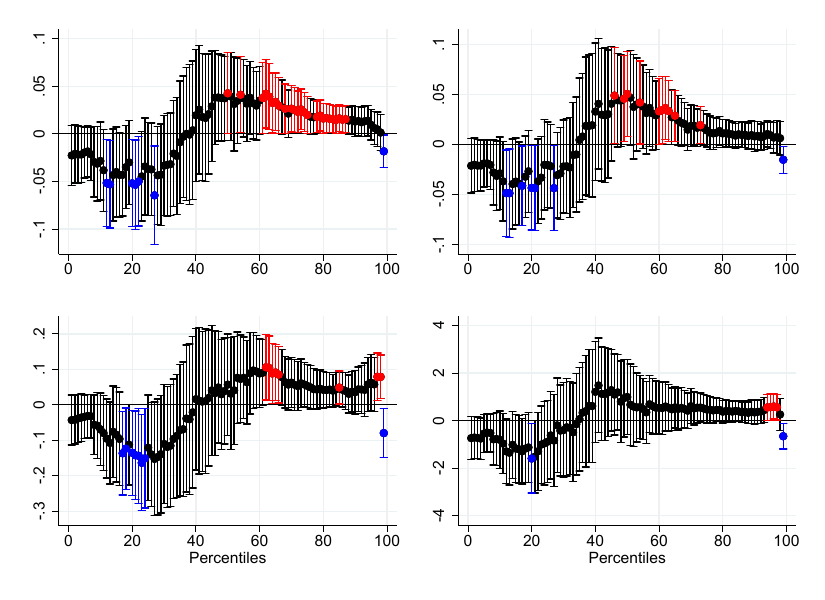}
\label{fig:perc_dem}
\begin{flushleft}
\footnotesize 
\singlespacing
\textit{Note}: In the first row, democracy is V-DEM Polyarchy, and V-DEM Liberal; in the second, democracy is Democracy-Dictatorship and Polity IV. Each graph shows $\beta$ coefficients for each percentile , with confidence intervals at the $90\%$ confidence level. In blue, percentiles with all the interval below zero; in red, above zero.    
\end{flushleft}
\end{figure}

Figure (\ref{fig:perc_dem}) presents the estimated coefficients for every percentile across the different measures of democracy.

The figure is highly illustrative of the association between income percentiles and democracy. In the lower part of the distribution, several percentiles display a negative relationship with democracy: the greater the income share of these poorer groups, the lower the democracy index, or the lower its likelihood. A similar pattern emerges at the top of the distribution. Regardless of the democracy measure used, the income share of the top 1 percent is consistently and negatively correlated with all of them.

In the middle of the distribution, in contrast, many percentiles exhibit a positive correlation between income share and democracy. In other words, a larger share accruing to the middle of the distribution is associated with higher levels of democracy. Interestingly, however, these percentiles are not symmetrically centered around the median. Instead, they are located closer to the upper end of the distribution. Across all four measures of democracy, the average percentile with a significant positive coefficient is 75; this falls slightly to 72 when computing the weighted average using only the positive coefficients. In any case, these values are clearly skewed toward the right of the distribution.

This paper shows that the income distribution can be divided into three groups based on their relationship to the distribution's overall inequality. Figure (\ref{fig:perc_dem}) complements this idea from a different angle. The estimates suggest that the relationship between income distribution with democracy also reveals the presence of three distinct groups within society.

\subsection{Middle Class and Democracy}

I now turn to the estimation of middle-class associations. The empirical specification corresponds to equation (\ref{eq:specification}) and uses several alternative measures of MIDCLASS. The set of controls is identical to that used in the previous section: the logarithm of GDP per capita, the Gini coefficient and its squared term, and average years of schooling. The fixed-effects estimator exploits within-country variation over time.

The regressions consider two broad types of middle-class definitions. The first corresponds to standard measures commonly used in the literature. Specifically, I include the $(20,80)$ middle class as defined by \citet{easterly2001middle}, and the $(75,200\%)$ middle class proposed by the OECD. The previous section discusses the evolution of both these measures. In addition, I consider the $(40,90)$ or Palma middle class \citep{palma2006globalizing}. 

The second type consists of the endogenous middle-class measures proposed in this paper. I consider middle classes of size $M=30$ and $M=50$, with global percentile ranges given by $(65,95)$ and $(48,98)$, respectively. I further include country-specific middle-class definitions, as discussed in subsection 5.5. I denoted these classes by their country-average initial and final percentiles - $(62.1,92.1)$ and $(46.7,96.7)$ - although the exact percentile thresholds vary by country.

\begin{figure}[H]
    \centering
\centering
  \caption{Middle Classes and Democracy}
  \includegraphics[width=1.0 \textwidth]{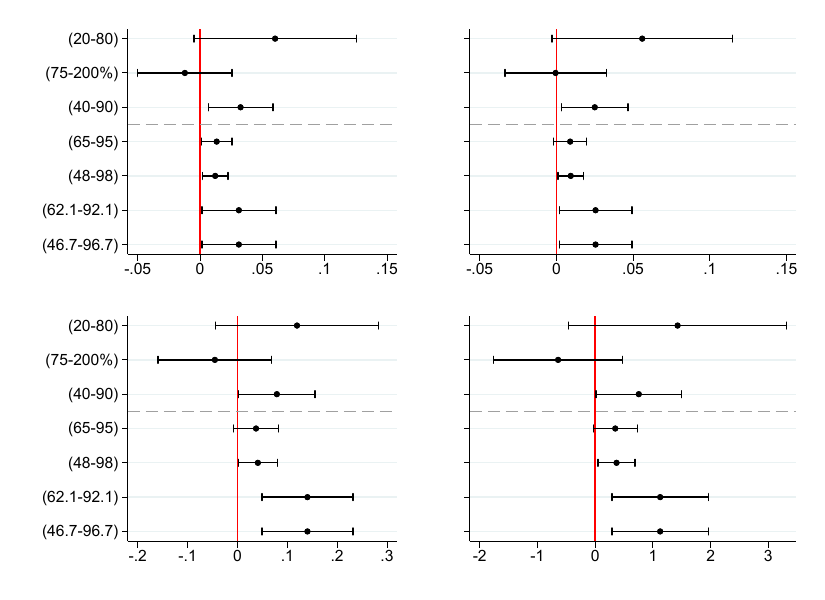}
\label{fig:mc_dem}
\begin{flushleft}
\footnotesize 
\singlespacing
\textit{Note}: In the first row, democracy is V-DEM Polyarchy, and V-DEM Liberal; in the second, democracy is Democracy-Dictatorship and Polity IV. The graphs show coefficients for each middle class, with confidence intervals at the $90\%$ confidence level. The dashed line separates middles classes: above  the line are the standard middle classes in the literature and below, the insensitive middle classes introduced in this paper.       
\end{flushleft}
\end{figure}

Figure (\ref{fig:mc_dem}) presents the results for each measure of democracy. The figure reports the estimated coefficients for each middle-class definition, along with 90\% confidence intervals.

In Figure (\ref{fig:mc_dem}), the standard middle-class coefficients are reported above the dashed line. The two conventional measures—the $(20,80)$ and $(75,200\%)$ middle classes—are not statistically significant for any of the democracy indicators. This result is interesting, given that these are among the most widely used definitions of the middle class in the economics literature. By contrast, the confidence interval for the Palma middle class lies slightly above zero in three of the four estimations. In other words, the correlation with democracy is stronger for a middle class tilted toward the upper part of the income distribution. 

The global middle-class measures, for both $M=30$ and $M=50$, lie at the margin of statistical significance at the $10\%$ level. For the $(65,95)$ class, significance is rejected in three out of four estimations, whereas the opposite pattern holds for the $(48,98)$ class. Overall, these global middle classes display positive but weak correlations with democracy.

The final two measures correspond to the country-specific middle classes. These classes are positively associated with democracy across all estimations. Statistical significance is achieved at the $10\%$ level for V-Dem and at the $5\%$ level for Dict-Dem and Polity, for both $M=30$ and $M=50$. In fact, the country-specific middle classes exhibit the strongest association with democracy among all the definitions considered in the figure.

The positive relationship between country-specific middle classes and democracy is reassuring for the political economy interpretation of social classes proposed in this paper. These middle classes correspond to the set of income percentiles that are insensitive to changes in inequality within a given country over time. In this sense, they are the groups with the least stake in distributive conflict. Accordingly, a positive relationship between the income share of these groups and democratic stability is theoretically expected. The empirical results point precisely in this direction.

\section{Conclusion}

Although income has a continuous distribution, much of the theoretical and empirical literature simplifies analysis by classifying individuals into discrete groups. When those groups or classes are defined endogenously, they typically rely on characteristics other than income—such as education, occupation, or productive factors. When income alone is used, by contrast, class boundaries are arbitrary.  

This paper proposes a definition of social classes based solely on the income distribution itself and on how different segments of that distribution respond to changes in inequality.  The poor are those whose share of total income declines as inequality rises; the rich are those whose share increases; and between them lies the middle class, defined as the group whose income share is invariant to changes in inequality. In this sense, income distribution is a “tale of three classes.”

Defining the middle class from a sample implies that its size for a country necessarily depends on other countries or the potential future trajectories within that country. Given that no definition based on a single income distribution is free from arbitrariness, this dependence is the unavoidable cost of endogenizing social classes. Crucially, however, the results are remarkably robust to the use of alternative samples: across countries and over time, the inequality-insensitive middle class consistently concentrates near the upper end of the income distribution. This empirical regularity suggests that the proposed definition captures a meaningful and stable economic group rather than a statistical artifact.

Admittedly, the choice of parameter $M$ - the size of the middle group - remains discretionary, affecting the size and composition of the other classes as well. Higher values of $M$ generate smaller poor and rich groups and increase polarization between them, lowering the median income of the poor while raising that of the rich. While this introduces discretion, it also allows for systematic exploration of how class structure varies with alternative assumptions about sizes, making these trade-offs explicit.

\bibliography{bib_middleclass}{}

@article{atkinson2013identification,
  title={On the identification of the middle class},
  author={Atkinson, Anthony B and Brandolini, Andrea},
  journal={Income inequality: Economic disparities and the middle class in affluent countries},
  pages={77--100},
  year={2013}
}

@article{lopez2014vulnerability,
  title={A vulnerability approach to the definition of the middle class},
  author={L{\'o}pez-Calva, Luis F and Ortiz-Juarez, Eduardo},
  journal={The Journal of Economic Inequality},
  volume={12},
  pages={23--47},
  year={2014},
  publisher={Springer}
}

@article{eisenhauer2008economic,
  title={An economic definition of the middle class},
  author={Eisenhauer, Joseph},
  booktitle={Forum for Social Economics},
  volume={37},
  number={2},
  pages={103--113},
  year={2008},
  organization={Taylor \& Francis}
}

@article{pressman2007decline,
  title={The decline of the middle class: an international perspective},
  author={Pressman, Steven},
  journal={Journal of Economic Issues},
  volume={41},
  number={1},
  pages={181--200},
  year={2007},
  publisher={Taylor \& Francis}
}

@article{banerjee2008middle,
  title={What is middle class about the middle classes around the world?},
  author={Banerjee, Abhijit V and Duflo, Esther},
  journal={Journal of economic perspectives},
  volume={22},
  number={2},
  pages={3--28},
  year={2008},
  publisher={American Economic Association}
}

@article{foster2010polarization,
  title={Polarization and the Decline of the Middle Class: Canada and the US},
  author={Foster, James E and Wolfson, Michael C},
  journal={The Journal of Economic Inequality},
  volume={8},
  pages={247--273},
  year={2010},
  publisher={Springer}
}

@article{easterly2001middle,
  title={The middle class consensus and economic development},
  author={Easterly, William},
  journal={Journal of economic growth},
  volume={6},
  number={4},
  pages={317--335},
  year={2001},
  publisher={Springer}
}

@article{barro1999determinants,
  title={Determinants of democracy},
  author={Barro, Robert J},
  journal={Journal of Political economy},
  volume={107},
  number={S6},
  pages={S158--S183},
  year={1999},
  publisher={The University of Chicago Press}
}

@article{murphy1989income,
  title={Income distribution, market size, and industrialization},
  author={Murphy, Kevin M and Shleifer, Andrei and Vishny, Robert},
  journal={The Quarterly Journal of Economics},
  volume={104},
  number={3},
  pages={537--564},
  year={1989},
  publisher={MIT Press}
}

@article{palma2011homogeneous,
  title={Homogeneous middles vs. heterogeneous tails, and the end of the ‘inverted-U’: It's all about the share of the rich},
  author={Palma, Jos{\'e} Gabriel},
  journal={Development and Change},
  volume={42},
  number={1},
  pages={87--153},
  year={2011},
  publisher={Wiley Online Library}
}

@article{palma2006globalizing,
  title={Globalizing inequality:‘centrifugal’and ‘centripetal’forces at work},
  author={Palma, Jos{\'e} Gabriel},
  year={2006},
  publisher={Citeseer}
}

@article{palma2019behind,
  title={Behind the Seven Veils of Inequality. What if it's all about the Struggle within just One Half of the Population over just One Half of the National Income?},
  author={Palma, Jos{\'e} Gabriel},
  journal={Development and Change},
  volume={50},
  number={5},
  pages={1133--1213},
  year={2019},
  publisher={Wiley Online Library}
}

@article{cobham2016inequality,
  title={Inequality and the tails: the Palma proposition and ratio},
  author={Cobham, Alex and Schl{\"o}gl, Lukas and Sumner, Andy},
  journal={Global Policy},
  volume={7},
  number={1},
  pages={25--36},
  year={2016},
  publisher={Wiley Online Library}
}

@article{krozer2015inequality,
  title={The inequality we want: How much is too much?},
  author={Krozer, Alice},
  journal={Journal of International Commerce, Economics and Policy},
  volume={6},
  number={03},
  pages={1550016},
  year={2015},
  publisher={World Scientific}
}

@article{benhabib2018skewed,
  title={Skewed wealth distributions: Theory and empirics},
  author={Benhabib, Jess and Bisin, Alberto},
  journal={Journal of Economic Literature},
  volume={56},
  number={4},
  pages={1261--1291},
  year={2018},
  publisher={American Economic Association 2014 Broadway, Suite 305, Nashville, TN 37203-2425}
}

@article{reed2004double,
  title={The double Pareto-lognormal distribution—a new parametric model for size distributions},
  author={Reed, William J and Jorgensen, Murray},
  journal={Communications in Statistics-Theory and Methods},
  volume={33},
  number={8},
  pages={1733--1753},
  year={2004},
  publisher={Taylor \& Francis}
}

@article{toda2012double,
  title={The double power law in income distribution: Explanations and evidence},
  author={Toda, Alexis Akira},
  journal={Journal of Economic Behavior \& Organization},
  volume={84},
  number={1},
  pages={364--381},
  year={2012},
  publisher={Elsevier}
}

@article{birdsall2010indispensable,
  title={The (indispensable) middle class in developing countries},
  author={Birdsall, Nancy},
  journal={Equity and growth in a globalizing world},
  volume={157},
  year={2010},
  publisher={Washington, DC: World Bank}
}

@article{cashell2008middle,
  title={Who are the “middle class”?},
  author={Cashell, Brian W},
  year={2008}
}

@article{ravallion2010developing,
  title={The developing world’s bulging (but vulnerable) middle class},
  author={Ravallion, Martin},
  journal={World development},
  volume={38},
  number={4},
  pages={445--454},
  year={2010},
  publisher={Elsevier}
}

@article{rasch2017measuring,
  title={Measuring the middle class in middle-income countries},
  author={Rasch, Rebecca},
  booktitle={Forum for Social Economics},
  volume={46},
  number={4},
  pages={321--336},
  year={2017},
  organization={Taylor \& Francis}
}

@article{milanovic2002decomposing,
  title={Decomposing world income distribution: Does the world have a middle class?},
  author={Milanovic, Branko and Yitzhaki, Shlomo},
  journal={Review of income and wealth},
  volume={48},
  number={2},
  pages={155--178},
  year={2002},
  publisher={Wiley Online Library}
}

@book{chancel2022world,
  title={World inequality report 2022},
  author={Chancel, Lucas and Piketty, Thomas and Saez, Emmanuel and Zucman, Gabriel},
  year={2022},
  publisher={Harvard University Press}
}

@article{alvaredo2017global,
  title={Global inequality dynamics: New findings from WID. world},
  author={Alvaredo, Facundo and Chancel, Lucas and Piketty, Thomas and Saez, Emmanuel and Zucman, Gabriel},
  journal={American Economic Review},
  volume={107},
  number={5},
  pages={404--409},
  year={2017},
  publisher={American Economic Association 2014 Broadway, Suite 305, Nashville, TN 37203}
}

@book{gornick2014income,
  title={Income inequality: Economic disparities and the middle class in affluent countries},
  author={Gornick, Janet C and J{\"a}ntti, Markus},
  year={2014},
  publisher={Stanford University Press}
}

@article{edo2021multidimensional,
  title={A multidimensional approach to measuring the middle class},
  author={Edo, Mar{\'\i}a and Escudero, Walter Sosa and Svarc, Marcela},
  journal={The Journal of Economic Inequality},
  volume={19},
  pages={139--162},
  year={2021},
  publisher={Springer}
}

@article{ricci2020measure,
  title={How to measure the middle class: approaches from economics},
  author={Ricci, Chiara Assunta},
  journal={DaSTU Working},
  year={2020}
}

@article{kharas2010new,
  title={The new global middle class: A cross-over from West to East},
  author={Kharas, Homi and Gertz, Geoffrey},
  journal={Wolfensohn Center for Development at Brookings},
  pages={1--14},
  year={2010}
}

@article{thurow1987surge,
  title={A surge in inequality},
  author={Thurow, Lester C},
  journal={Scientific American},
  volume={256},
  number={5},
  pages={30--37},
  year={1987},
  publisher={JSTOR}
}

@article{levy1987middle,
  title={The middle class: is it really vanishing?},
  author={Levy, Frank},
  journal={The Brookings Review},
  volume={5},
  number={3},
  pages={17--21},
  year={1987},
  publisher={JSTOR}
}

@article{reeves2018dozen,
  title={A dozen ways to be middle class},
  author={Reeves, Richard V and Guyot, Katherine and Krause, Eleanor},
  journal={The Brookings Institute. Retrieved May},
  volume={18},
  pages={2018},
  year={2018}
}

@article{benhabib2006political,
  title={The political economy of redistribution under democracy},
  author={Benhabib, Jess and Przeworski, Adam},
  journal={Economic Theory},
  volume={29},
  pages={271--290},
  year={2006},
  publisher={Springer}
}

@article{acemoglu1997prometheus,
  title={Was Prometheus unbound by chance? Risk, diversification, and growth},
  author={Acemoglu, Daron and Zilibotti, Fabrizio},
  journal={Journal of political economy},
  volume={105},
  number={4},
  pages={709--751},
  year={1997},
  publisher={The University of Chicago Press}
}

@article{galor1997technological,
  title={Technological progress, mobility, and economic growth},
  author={Galor, Oded and Tsiddon, Daniel},
  journal={The American Economic Review},
  pages={363--382},
  year={1997},
  publisher={JSTOR}
}

@article{deininger1996new,
  title={A new data set measuring income inequality},
  author={Deininger, Klaus and Squire, Lyn},
  journal={The World Bank Economic Review},
  volume={10},
  number={3},
  pages={565--591},
  year={1996},
  publisher={Oxford University Press}
}

@article{alesina1996income,
  title={Income distribution, political instability, and investment},
  author={Alesina, Alberto and Perotti, Roberto},
  journal={European economic review},
  volume={40},
  number={6},
  pages={1203--1228},
  year={1996},
  publisher={Elsevier}
}

@article{piketty2006evolution,
  title={The evolution of top incomes: a historical and international perspective},
  author={Piketty, Thomas and Saez, Emmanuel},
  journal={American economic review},
  volume={96},
  number={2},
  pages={200--205},
  year={2006},
  publisher={American Economic Association}
}

@book{piketty2014capital,
  title={Capital in the twenty-first century},
  author={Piketty, Thomas},
  year={2014},
  publisher={Harvard University Press}
}

@article{deininger1998new,
  title={New ways of looking at old issues: inequality and growth},
  author={Deininger, Klaus and Squire, Lyn},
  journal={Journal of development economics},
  volume={57},
  number={2},
  pages={259--287},
  year={1998},
  publisher={Elsevier}
}

@article{milanovic10still,
  title={Why I am still not excited by the Palma index},
  author={Milanovic, Branko},
  journal={globalinequality (blog)},
  year={2015},
  volume={10}
}

@article{nolan2011economic,
  title={Economic inequality, poverty, and social exclusion},
  author={Nolan, Brian and Ive, Marx},
  year={2011}
}

@article{coppedge2011conceptualizing,
  title={Conceptualizing and measuring democracy: A new approach},
  author={Coppedge, Michael and Gerring, John and Altman, David and Bernhard, Michael and Fish, Steven and Hicken, Allen and Kroenig, Matthew and Lindberg, Staffan I and McMann, Kelly and Paxton, Pamela and others},
  journal={Perspectives on politics},
  volume={9},
  number={2},
  pages={247--267},
  year={2011},
  publisher={Cambridge University Press}
}

@article{coppedge2019methodology,
  title={The methodology of “varieties of democracy”(V-Dem)},
  author={Coppedge, Michael and Gerring, John and Knutsen, Carl Henrik and Krusell, Joshua and Medzihorsky, Juraj and Pernes, Josefine and Skaaning, Svend-Erik and Stepanova, Natalia and Teorell, Jan and Tzelgov, Eitan and others},
  journal={Bulletin of Sociological Methodology/Bulletin de M{\'e}thodologie Sociologique},
  volume={143},
  number={1},
  pages={107--133},
  year={2019},
  publisher={Sage Publications Sage UK: London, England}
}

@article{coppedge2008two,
  title={Two persistent dimensions of democracy: Contestation and inclusiveness},
  author={Coppedge, Michael and Alvarez, Angel and Maldonado, Claudia},
  journal={The Journal of Politics},
  volume={70},
  number={3},
  pages={632--647},
  year={2008},
  publisher={Cambridge University Press New York, USA}
}

@article{munck2002conceptualizing,
  title={Conceptualizing and measuring democracy: Evaluating alternative indices},
  author={Munck, Gerardo L and Verkuilen, Jay},
  journal={Comparative political studies},
  volume={35},
  number={1},
  pages={5--34},
  year={2002},
  publisher={Sage Publications Sage CA: Thousand Oaks, CA}
}

@article{boese2019not,
  title={How (not) to measure democracy},
  author={Boese, Vanessa A},
  journal={International Area Studies Review},
  volume={22},
  number={2},
  pages={95--127},
  year={2019},
  publisher={SAGE Publications Sage UK: London, England}
}

@book{przeworski2000democracy,
  title={Democracy and development: Political institutions and well-being in the world, 1950-1990},
  author={Przeworski, Adam},
  year={2000},
  publisher={Cambridge University Press}
}

@book{dahl1971polyarchy,
  title={Polyarchy: Participation and opposition},
  author={Dahl, Robert A},
  year={1971},
  publisher={Yale university press}
}

@article{alvarez1996classifying,
  title={Classifying political regimes},
  author={Alvarez, Mike and Cheibub, Jos{\'e} Antonio and Limongi, Fernando and Przeworski, Adam},
  journal={Studies in comparative international development},
  volume={31},
  number={2},
  pages={3--36},
  year={1996},
  publisher={Springer}
}

@article{cheibub2010democracy,
  title={Democracy and dictatorship revisited},
  author={Cheibub, Jos{\'e} Antonio and Gandhi, Jennifer and Vreeland, James Raymond},
  journal={Public choice},
  volume={143},
  number={1},
  pages={67--101},
  year={2010},
  publisher={Springer}
}

@article{bjornskov2020regime,
  title={Regime types and regime change: A new dataset on democracy, coups, and political institutions},
  author={Bj{\o}rnskov, Christian and Rode, Martin},
  journal={The Review of International Organizations},
  volume={15},
  number={2},
  pages={531--551},
  year={2020},
  publisher={Springer}
}

@article{marshall2020polity5,
  title={Polity5: Political regime characteristics and transitions, 1800--2018},
  author={Marshall, Monty G and Gurr, Ted Robert},
  journal={Center for Systemic Peace},
  volume={2},
  year={2020}
}

@book{AristotlePolitics,
  author    = {Aristotle},
  title     = {Politics},
  translator = {Carnes Lord},
  edition   = {2nd},
  publisher = {University of Chicago Press},
  year      = {2013},
}

@article{lipset1959some,
  title={Some social requisites of democracy: Economic development and political legitimacy1},
  author={Lipset, Seymour Martin},
  journal={American political science review},
  volume={53},
  number={1},
  pages={69--105},
  year={1959},
  publisher={Cambridge University Press}
}

@book{moore1966social,
  title={Social origins of dictatorship and democracy: Lord and peasant in the making of the modern world},
  author={Moore, Barrington},
  year={1966},
  publisher={Beacon Press}
}

@article{huber1993impact,
  title={The impact of economic development on democracy},
  author={Huber, Evelyne and Rueschemeyer, Dietrich and Stephens, John D},
  journal={Journal of economic perspectives},
  volume={7},
  number={3},
  pages={71--86},
  year={1993},
  publisher={American Economic Association}
}

@article{acemoglu2008income,
  title={Income and democracy},
  author={Acemoglu, Daron and Johnson, Simon and Robinson, James A and Yared, Pierre},
  journal={American economic review},
  volume={98},
  number={3},
  pages={808--842},
  year={2008},
  publisher={American Economic Association}
}

@book{boix2003democracy,
  title={Democracy and redistribution},
  author={Boix, Carles},
  year={2003},
  publisher={Cambridge University Press}
}

@book{acemoglu2006economic,
  title={Economic origins of dictatorship and democracy},
  author={Acemoglu, Daron and Robinson, James A},
  year={2006},
  publisher={Cambridge university press}
}

@article{benhabib2013income,
  title={Income and democracy: Evidence from nonlinear estimations},
  author={Benhabib, Jess and Corvalan, Alejandro and Spiegel, Mark M},
  journal={Economics Letters},
  volume={118},
  number={3},
  pages={489--492},
  year={2013},
  publisher={Elsevier}
}

\end{document}